\documentclass[a4paper,11pt]{article}
\pdfoutput=1
\usepackage{jheppub}
\usepackage{amsmath,amssymb,latexsym}
\usepackage{bm}
\usepackage{hepunits}
\usepackage[svgnames]{xcolor}

\usepackage{graphicx}
\usepackage{float}
\usepackage{subfigure}

\usepackage{mathtools}
\usepackage{braket}

\usepackage{natbib}
\usepackage{graphicx}

\usepackage{shuffle} 
\usepackage{slashed} 

\usepackage{cprotect}
\usepackage{multirow}
\usepackage{diagbox}

\newcommand{\zcal}{\mathcal{Z}}
\newcommand{\scal}{\mathcal{S}}

\DeclareUnicodeCharacter{2212}{-}
\DeclareUnicodeCharacter{2217}{*}

\newcommand{\nn}{\nonumber}

\newcommand{\mC}{\mathcal}

\def \lt {\left}
\def \rt {\right}
\def \eps {\epsilon}
\def \veps {\varepsilon}
\def \fc{\frac}
\def \td{\tilde}

\allowdisplaybreaks

\title{Two-loop QCD amplitudes for $t\bar{t}H$ production from boosted limit}

\author[a,b]{Guoxing Wang,}
\author[a]{Tianya Xia,}
\author[a]{Li Lin Yang,}
\author[c]{Xiaoping Ye}
\affiliation[a]{Zhejiang Institute of Modern Physics, School of Physics, Zhejiang University, Hangzhou 310027, China}
\affiliation[b]{Laboratoire de Physique Th\'eorique et Hautes Energies (LPTHE), UMR 7589, Sorbonne Universit\'e et CNRS, 4 place Jussieu, 75252 Paris Cedex 05, France}
\affiliation[c]{School of Physics, Peking University, Beijing 100871, China}
\emailAdd{wangguoxing2015@pku.edu.cn}
\emailAdd{xiatianya@zju.edu.cn}
\emailAdd{yanglilin@zju.edu.cn}
\emailAdd{yexiaoping@pku.edu.cn}

\abstract{
The production of a Higgs boson in association with a top-antitop quark pair ($t\bar{t}H$) holds significant importance in directly probing the top-quark Yukawa coupling, which is related to various fundamental questions in high energy physics.
This paper focuses on the calculation of two-loop amplitudes for $t\bar t H$ production at hadron colliders in the high-energy boosted limit. The calculation employs our recently developed mass-factorization formula. 
To validate the accuracy of our approximate methods, we compare our results for the one-loop amplitudes and the two-loop infrared poles with the exact calculations.
We then provide predictions for the finite parts of the two-loop amplitudes. By combining the contributions from real emissions, our results can be utilized to compute the next-to-next-to-leading order differential cross sections for $t\bar t H$ production in the high-energy boosted limit.
}

\begin{document}

\maketitle

\clearpage

\section{Introduction}

Since the discovery of the Higgs boson by the ATLAS and CMS collaborations in 2012 \cite{ATLAS:2012yve, CMS:2012qbp}, the precision test of the properties of the Higgs boson has become one of the most important topics in high-energy physics. In particular, the Yukawa coupling between the top quark and the Higgs boson plays an important role in pursuing the answers to many deep questions, such as the origin of the masses of fundamental fermions, the stability of the electroweak vacuum, and the matter-anti-matter asymmetry in our observable universe. 
The ATLAS and CMS collaborations at the Large Hadron Collider (LHC) have observed the Higgs production associated with a top-antitop quark pair ($t\bar{t}H$ production) in 2018 \cite{ATLAS:2018mme, CMS:2018uxb}. This allows us to directly probe the top-quark Yukawa coupling. The possible violation of the Charge-Parity (CP) symmetry in the coupling has also been investigated experimentally \cite{CMS:2020cga, ATLAS:2020ior}.
With the accumulation of data in the near future, the experimental accuracy for the cross section of this process can reach $2\%$ at the High-Luminosity LHC (HL-LHC) \cite{Cepeda:2019klc}. To match the expected precision of the HL-LHC, it is highly desirable to have equally accurate theoretical predictions.

The leading-order (LO) cross sections for $t\bar{t}H$ production were present in \cite{Ng:1983jm, Kunszt:1984ri}, and the next-to-leading order (NLO) QCD corrections have been calculated 20 years ago \cite{Beenakker:2001rj, Beenakker:2002nc, Reina:2001sf, Reina:2001bc, Dawson:2002tg, Genovese:2003dp}. The NLO eletroweak (EW) corrections were reported in \cite{Frixione:2014qaa, Zhang:2014gcy, Frixione:2015zaa}. The resummation of soft-gluon contributions close to the partonic threshold up to NNLL was considered in \cite{Kulesza:2015vda, Broggio:2015lya, Broggio:2016lfj, Kulesza:2017ukk, Broggio:2019ewu, Kulesza:2020nfh}. The mixed QCD+EW corrections were calculated in \cite{Denner:2016wet}. Beyond all these, the next-to-next-to-leading order (NNLO) QCD corrections are the current frontier. In the NNLO corrections, in addition to the two-loop amplitudes, the higher order contributions of one-loop amplitudes in the dimensional regulator $\epsilon$ are also necessary. In \cite{Chen:2022nxt}, the $\mathcal{O}(\epsilon^1)$ contributions for both the quark-antiquark annihilation channel and the gluon fusion channel were obtained and used to extract the Infrared (IR) singularities of corresponding two-loop amplitudes. The $\mathcal{O}(\epsilon^2)$ contributions for the gluon fusion channel have recently been known in \cite{Buccioni:2023okz}. The contributions from off-diagonal partonic channels are present in \cite{Catani:2021cbl}, which turn out to be at the per-mill level.
In \cite{Catani:2022mfv}, the so-called soft Higgs approximation was applied to calculate the most difficult two-loop amplitudes. This approximation is valid in the limit where the momentum of the Higgs boson is small compared to other scales in the process. Such an approximation turns out to be reasonable for calculating the total cross sections, but may not be applicable to more exclusive differential cross sections. On the other hand, the two-loop master integrals for leading-color QCD scattering amplitudes with a closed light-quark loop in $t\bar{t}H$ production were calculated in \cite{FebresCordero:2023gjh}, while the exact calculation of the two-loop amplitudes is still beyond the reach of current methods.

In this paper, we take a kinematic limit different from that in \cite{Catani:2022mfv}. We consider the two-loop amplitudes for $t\bar{t}H$ production in the high-energy boosted limit, where the momentum-invariants are much larger than the mass of the top quark, i.e., $|s_{ij}|\gg m_t^2$ for all external legs $i$ and $j$. In this limit, a general mass-factorization formula was given in \cite{Mitov:2006xs, Becher:2007cu, Engel:2018fsb, Wang:2023qbf}. Using this factorization formula, our main task becomes the calculation of the two-loop massless amplitudes, where $m_t$ and $m_H$ are set to zero. This is much simpler than the calculation of the fully massive amplitudes, and is made possible by the recent progresses in the integration-by-parts (IBP) reduction techniques \cite{Bohm:2018bdy, Bendle:2019csk, Boehm:2020ijp, Bendle:2021ueg, Wu:2023upw, Guan:2019bcx, Heller:2021qkz} and the results for the two-loop five-point master integrals \cite{Gehrmann:2015bfy, Papadopoulos:2015jft, Gehrmann:2018yef, Chicherin:2018mue, Abreu:2018rcw, Abreu:2018aqd, Chicherin:2018old, Badger:2019djh}.

The paper is organized as follows. In Sec.~\ref{sec:convention}, we give our notations and the definitions of various quantities used in our calculation. We introduce our approximate formula for the massive amplitudes in the high-energy boosted limit in Sec.~\ref{sec:amp}, and discuss the conventions for dealing with power-suppressed contributions in Sec.~\ref{sec:squaredamp}. We then present our numeric results in Sec.~\ref{sec:numeric} and conclude in Sec.~\ref{sec:conclusion}. We collect some length expressions in the appendices.

\section{Notations and definition}\label{sec:convention}
For the production of a Higgs boson associated with a top-antitop quark pair, we consider the following two partonic processes
\begin{align}
q_\beta(p_1)+\bar{q}_\alpha(p_2) &\to t_k(p_3)+\bar{t}_l(p_4)+H(p_5) \,, \\
g_a(p_1)+g_b(p_2) &\to t_k(p_3)+\bar{t}_l(p_4)+H(p_5) \,,
\end{align}
where $\alpha, \beta, a, b, k, l$ are color indices and $p_i$ the momenta of the external partons with $p_1^2=p_2^2=0$, $p_3^2=p_4^2=m_t^2$ and $p_5^5=m_H^2$. The kinematic variables are defined as $s_{ij}=(p_i + \sigma_{ij} p_j)^2$, where $\sigma_{ij}=+1$ if the momenta $p_i$ and $p_j$ are both incoming or outgoing, and $\sigma_{ij}=-1$ otherwise.

To facilitate the calculation of the amplitudes $\ket{\mC{M}_{q,g}}$, we decompose them in terms of color and spin (Lorentz) structures in the color~$\otimes$~spin space of external partons. Note that the subscript $q$ or $g$ specifies the quark-antiquark annihilation channel or the gluon fusion channel, respectively. The amplitudes $\ket{\mC{M}_{q,g}}$ can be expanded in the strong coupling constant $\alpha_s$ up to NNLO as
\begin{equation}
\ket{\mC{M}_{q,g}}=4\pi\alpha_s\lt[\ket{\mC{M}_{q,g}^{(0)}}+\fc{\alpha_s}{4\pi}\ket{\mC{M}_{q,g}^{(1)}}+\lt(\fc{\alpha_s}{4\pi}\rt)^2\ket{\mC{M}_{q,g}^{(2)}}+ \mC{O}(\alpha_s)^3 \rt]\,.
\end{equation}
The $l$-loop coefficient $\ket{\mC{M}^{(l)}_{q,g}}$ can be further decomposed in the color~$\otimes$~spin space
\begin{equation}
\ket{\mC{M}_{q,g}^{(l)}}=\sum_{I,i}c^{(l)q,g}_{Ii}\ket{c_I^{q,g}} \,\otimes \, \ket{d_i^{q,g}} \,,
\end{equation}
where $\ket{c_I^{q,g}}$ are the orthogonal color bases for the quark-antiquark annihilation channel and the gluon fusion channel, $\ket{d_i^{q,g}}$ denote independent spin structures. We employ the color-space formalism \cite{Catani:1996jh, Catani:1996vz}, and choose the color bases as
\begin{align}\label{eq:colorbasis}
   \ket{c^q_1} = \delta_{\alpha\beta}\,\delta_{kl} \,, &
    \quad 
   \ket{c^q_2} = (t^a)_{\alpha\beta}\,(t^a)_{kl} \,, \nn \\
   \ket{c^g_1} = \delta^{ab}\,\delta_{kl} \,,
    \quad 
   \ket{c^g_2} = i&f^{abc}\,(t^c)_{kl} \,,
    \quad 
   \ket{c^g_3} = d^{abc}\,(t^c)_{kl} \,.
\end{align}
To define the spin structures $\ket{d_i^{q,g}}$, we choose the conventional dimensional regularization (CDR) scheme and assume that the external partons live in $d$ spacetime dimensions. There are $28$ independent structures in the quark-antiquark annihilation channel and $40$ independent structures in the gluon fusion channel, which are given in Appendix~\ref{sec:massivebases}.

The ultraviolet (UV) divergences in the bare amplitudes $\ket{\mC{M}^{\rm bare}_{q,g}}$ are renormalized according to 
\begin{align}
\Ket{{\cal M}^R_{q,g}(\alpha_s, g_Y, m_t, \mu, \epsilon)} = \left(\frac{\mu^2 e^{\gamma_E}}{4\pi}\right)^{-3\epsilon/2} Z_{q,g} Z_Q \Ket{{\cal M}^{\text{bare}}_{q,g}(\alpha_s^0, g_Y^0, m_t^0,\epsilon)} ,
\label{eq:UVren}
\end{align}
where $Z_g$, $Z_q$ and $Z_Q$ are the on-shell wave-function renormalization constants for gluons, light- and heavy-quarks, respectively. We have suppressed the dependence of the amplitudes on other kinematic variables. The Yukawa coupling $g_Y$ is defined as
\begin{equation}\label{eq:yukawa}
g_Y = \frac{e \, m_t}{2m_W\sin(\theta_W)} \, .
\end{equation}
We renormalize the top-quark mass in the on-shell scheme: $m_t^0 = Z_m m_t$, and the Yukawa coupling is renormalized accordingly. The strong coupling constant $\alpha_s$ is renormalized in the $\overline{\text{MS}}$ scheme with $n_f=n_l+n_h$ active flavors. The relations between the bare couplings and the renormalized ones are given by
\begin{equation}\label{eq:renorcoupling}
\alpha_s^0 = \left( \frac{\mu^2 e^{\gamma_E}}{4\pi} \right)^{\epsilon} Z_{\alpha_s} \alpha_s \, , \quad g_Y^0 = \left( \frac{\mu^2 e^{\gamma_E}}{4\pi} \right)^{\epsilon/2} Z_m \, g_Y \, .
\end{equation}
The renormalization constants up to NNLO are given in the Appendix~\ref{sec:renormalconstant}. 

The renormalized form factors $c^{R;q,g}_{Ii}$ can be extracted as a linear combination of scalar Feynman integrals according to 
\begin{align}\label{eq:formfactorR}
 c^{R;q,g}_{Ii}=\sum_{j}\frac{\lt(\bm{D}_{q,g}^{-1}\rt)_{ij}}{\braket{c^{q,g}_I | c^{q,g}_I}}\lt[\bra{d^{q,g}_j}\otimes\braket{c^{q,g}_I | \mC{M}^{R}_{q,g}}\rt]\,,
\end{align}
where $\bm{D}_{q,g}$ are matrices in the space of spin structures, whose elements are defined by $\bm{D}^{q,g}_{ij} = \braket{d^{q,g}_i | d^{q,g}_j}$. The polarization sum of the two initial gluons yields
\begin{align}
\sum_s\veps^\mu(p_1,s)\veps^\nu(p_1,s)=\sum_s\veps^\mu(p_2,s)\veps^\nu(p_2,s)=-g^{\mu\nu}+\fc{p_1^\mu p_2^\nu+p_1^\nu p_2^\mu}{p_1\cdot p_2} \,.
\end{align}

The two-loop renormalized amplitudes $\ket{\mC{M}^{(2)R}_{q,g}}$ are the main object of study in this work. Note that the IR singularities of $\ket{\mC{M}^{(2)R}_{q,g}}$ have been presented in \cite{Chen:2022nxt}. The IR poles are obtained from the one-loop amplitude up to $\mathcal{O}(\epsilon^1)$ and a universal anomalous dimension matrix. However, for the finite part, one needs to calculate the highly non-trivial two-loop five-point Feynman integrals involving 7 physical scales. In this work, we explore the high-energy boosted limit, and use the mass-factorization \cite{Mitov:2006xs} to approximately calculate these two-loop amplitudes.

\section{Massive amplitudes in the high-energy boosted limit}\label{sec:amp}

\subsection{The massive amplitudes from mass-factorization}
\label{sec:mass-fac}

Consider the amplitude for a generic $2\to n$ partonic scattering process in QCD:
\begin{equation}
h_1(p_1,m_1)+h_2(p_2,m_2)\to h_3(p_3,m_3)+h_4(p_4,m_4)+\cdots+h_{n+2}(p_{n+2},m_{n+2}) \,,
\end{equation}
where $m_i$ is the mass of the parton $h_i$. We work in the high-energy boosted limit, where $|s_{ij}| \gg m_k^2$ for arbitrary $i,j,k$ and $i\neq j$. In this limit, the renormalized massive amplitude $\mC{M}^R$ can be factorized as \cite{Mitov:2006xs, Becher:2007cu, Engel:2018fsb, Wang:2023qbf}
\begin{multline}
\label{eq:QCDfac}
\Ket{\mC{M}^R\lt(\epsilon,\{p\},\{m\},\mu\rt)} = \prod_j \left(\zcal_{[j]}^{(m|0)}(\epsilon,\{m\},\mu)\right)^{1/2}
\\
\times \bm{\scal}(\epsilon,\{\tilde{p}\},\{m\},\mu) \Ket{\td{\mC{M}}^R\lt(\epsilon,\{\tilde{p}\},\mu\rt)} + \mC{O}(m^2) \,,
\end{multline}
where $\tilde{\mathcal{M}}^R$ is the renormalized massless amplitude with all parton masses taken to zero. We use $\{p\}$  and $\{\tilde{p}\}$ to collectively denote the sets of all external momenta for the massive and massless amplitudes, and $\{m\}$ to denote the set of all parton masses. The $\zcal$-factors $\zcal_{[j]}^{(m|0)}$ and the soft function $\bm{\scal}$ contain all the mass logarithms $\ln(\mu^2/m^2)$, where $\bm{\scal}$ only receives contributions from closed-loops of massive partons starting from $\mC{O}(\alpha_s^2)$. The expressions of $\zcal_{[j]}^{(m|0)}$ for gluons and quarks up to NNLO can be found in \cite{Mitov:2006xs, Czakon:2007ej, Czakon:2007wk}, up to terms proportional to $n_h^1 n_l^0$. These missing terms as well as the soft function can be found in \cite{Becher:2007cu, Engel:2018fsb, Wang:2023qbf}. For completeness, we give the explicit expressions in Appendix~\ref{sec:mczfac}.
Note that the factorization in \eqref{eq:QCDfac} is correct up to power corrections of $\mathcal{O}(m^2)$. These sub-leading contributions are currently under active investigations \cite{Gervais:2017yxv, Laenen:2020nrt, Fadin:2023phc}. In this work, we will keep only the leading power (LP) terms and neglect the power corrections.

For the $t\bar{t}H$ production process considered in this work, there is a massive Higgs boson in the final state in addition to massive top (anti-)quarks. The Higgs boson is color neutral and is not involved in QCD interactions. We can in principle keep the Higgs mass $m_H$ in $\tilde{\mathcal{M}}^R$ of Eq.~\eqref{eq:QCDfac}. However, since $m_H < m_t$, it is reasonable to expand in $m_H$ as well. Therefore, for simplicity, we choose to take $m_H \to 0$ in $\tilde{\mathcal{M}}^R$. Note that this does not introduce logarithmic terms of $m_H$.

We are now ready to define an approximate amplitudes for $t\bar{t}H$ production in the boosted limit, given by
\begin{multline}
\label{eq:QCDfac1}
\Ket{\bar{\mC{M}}^R_{q,g}\lt(\epsilon,\{p\},m_t,m_H,\mu\rt)} = \zcal_{[q,g]}^{(m|0)}(\epsilon,m_t,\mu) \, \zcal_{[t]}^{(m|0)}(\epsilon,m_t,\mu)
\\
\times \bm{\scal}(\epsilon,\{\tilde{p}\},m_t,\mu)\Ket{\td{\mC{M}}^R_{q,g}\lt(\epsilon,\{\tilde{p}\},\mu\rt)} \,.
\end{multline}
The amplitude $\bar{\mC{M}}^R$ provides an approximation to the full amplitude $\mC{M}^R$ up to power corrections in $m_t^2$ and $m_H^2$. The momenta in $\tilde{\mathcal{M}}^R$ satisfy the on-shell conditions $\tilde{p}_i^2 = 0$, and we define $\tilde{s}_{ij} \equiv \sigma_{ij} \, \tilde{p}_i \cdot \tilde{p}_j$. There are some degrees of freedom in choosing how $\{\tilde{p}\}$ are related to the original momenta $\{p\}$ in the massive amplitude. We will discuss our choice in the next Section. Before that, we turn to the actual calculation of the massless amplitude $\tilde{\mathcal{M}}^R$ at NNLO in the following.

\subsection{Calculation of the massless amplitudes}

Similar to the massive case, we renormalize the massless amplitudes according to 
\begin{align}
\Ket{{\td{\mC{M}}}^R_{q,g}(\alpha_s, g_Y, \mu, \epsilon)} = \left(\frac{\mu^2 e^{\gamma_E}}{4\pi}\right)^{-3\epsilon/2} \Ket{\td{\mC{M}}^{\text{bare}}_{q,g}(\alpha_s^0, g_Y^0, \epsilon)} \,,
\end{align}
where we have suppressed the dependence of the amplitudes on the kinematic variables.
Note that the on-shell wave-function renormalization constants for gluons and quarks are always unity in the massless case. The renormalization of the strong coupling constant $\alpha_s$ and the Yukawa coupling constant $g_Y$ is the same as Eq.~\eqref{eq:renorcoupling}.
The renormalized massless amplitudes can then be decomposed in the color $\otimes$ spin space into 
\begin{equation}\label{eq:decompsemassless}
\ket{\td{\mC{M}}_{q,g}^R}=\sum_{I,i}\td{c}^{R;q,g}_{Ii}\ket{c_I^{q,g}} \,\otimes \, \ket{\td{d}_i^{q,g}} \,,
\end{equation}
where the color basis is the same as Eq.~\eqref{eq:colorbasis}.
However, the number of linearly independent spin (Lorentz) structures $\ket{\td{d}_i^{q,g}}$ in the massless case is different from that in the massive case. There are $14$ independent structures for the quark-antiquark annihilation channel and $20$ independent structures for the gluon fusion channel. We list them in Appendix~\ref{sec:masslessbases}. Note that the number of massless spin structures is exactly one-half of that of massive ones. This is due to the chirality conservation of perturbative QCD vertices, which implies that the chirality flipping terms vanish when taking $m_t=0$.

\begin{figure}[t]
\centering  
\subfigure[planar pentagon-box (PB)]{
\label{Fig.PB}
\includegraphics[width=0.35\textwidth]{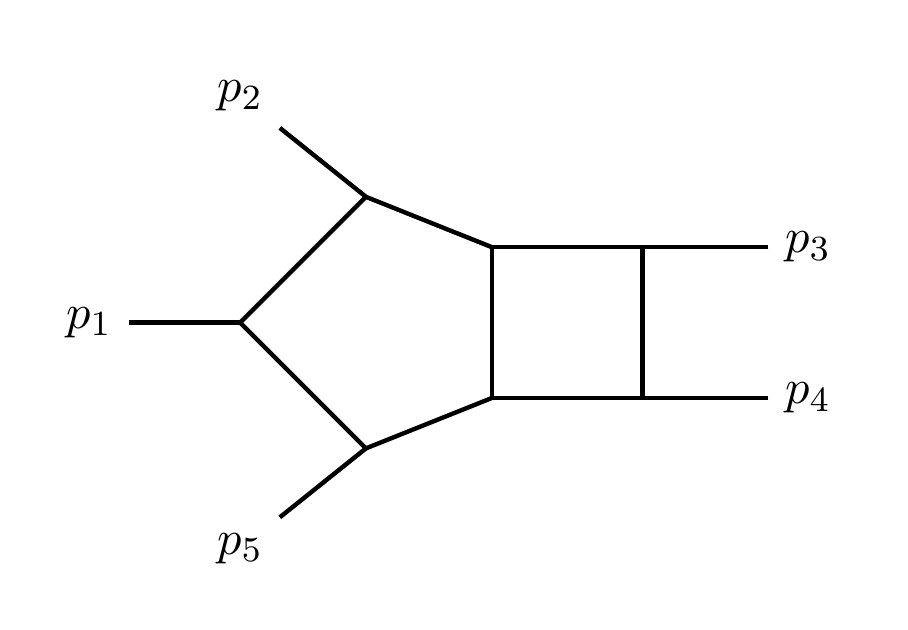}}
\subfigure[non-planar hexagon-box (HB)]{
\label{Fig.HB}
\includegraphics[width=0.35\textwidth]{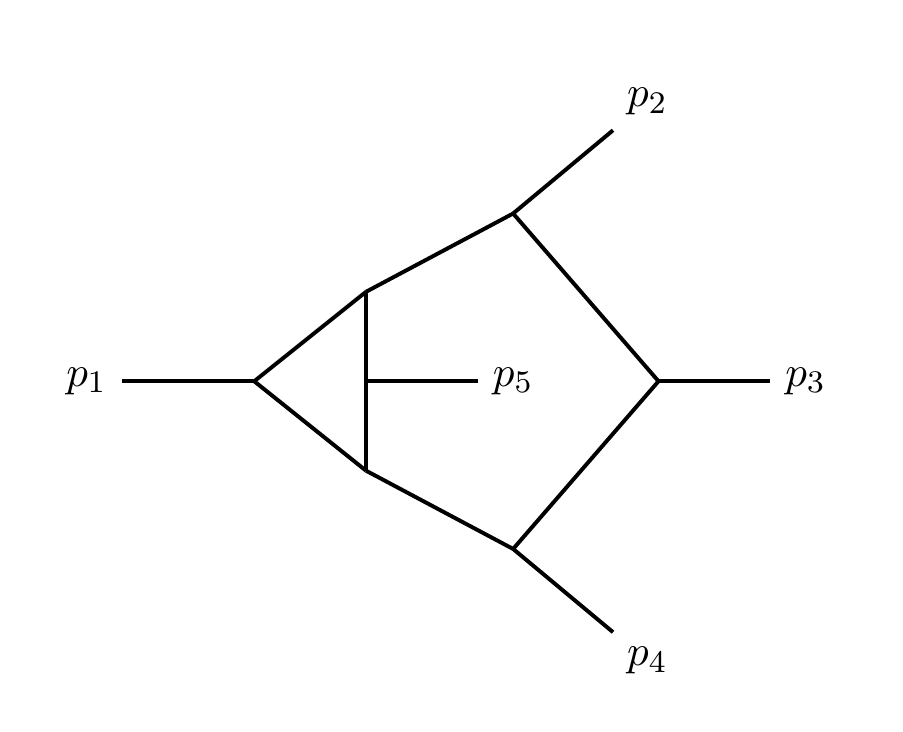}}
\subfigure[non-planar double pentagon (DP)]{
\label{Fig.DP}
\includegraphics[width=0.35\textwidth]{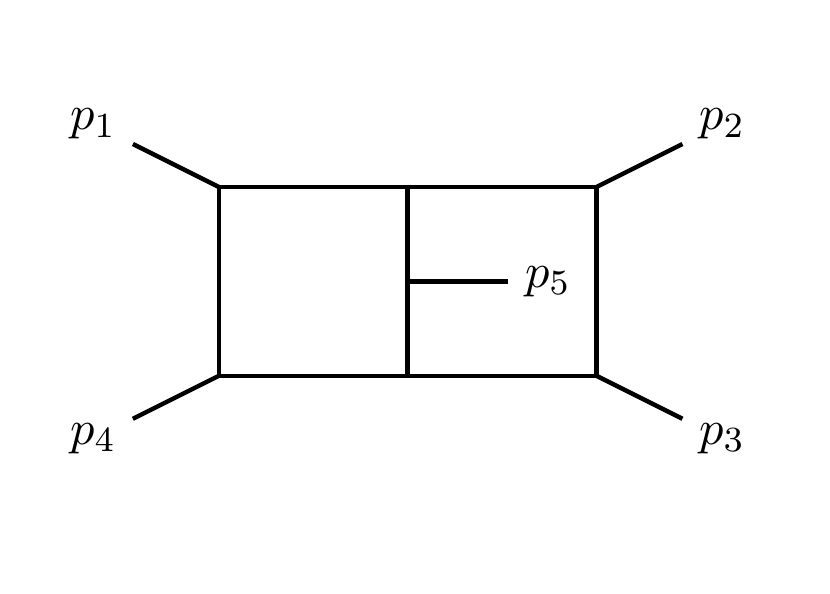}}
\subfigure[planar hexagon-triangle (HT)]{
\label{Fig.HT}
\includegraphics[width=0.35\textwidth]{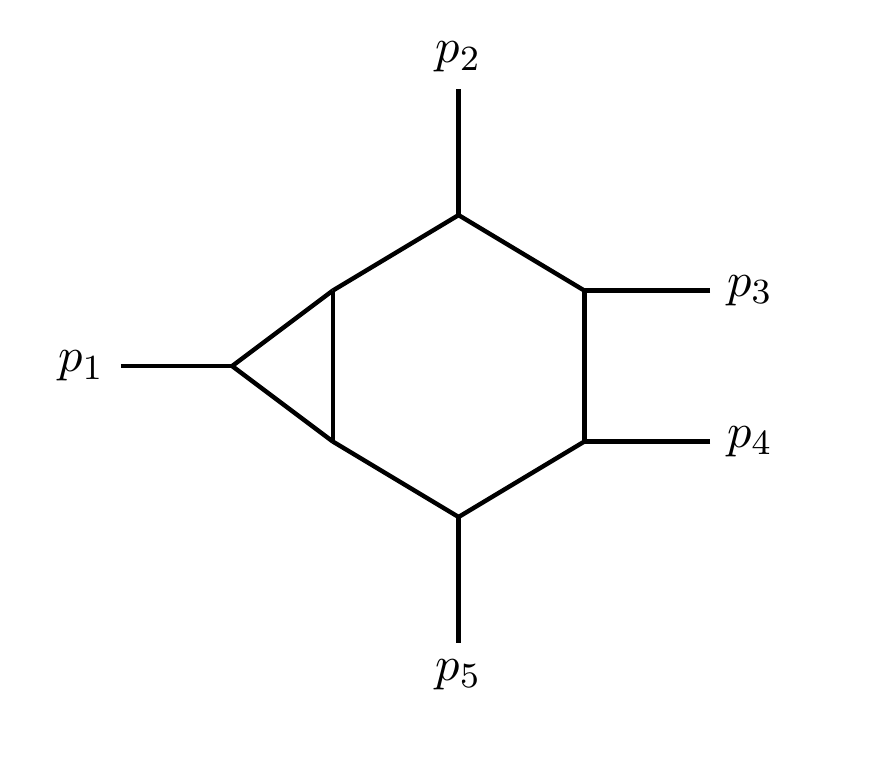}}
\caption{Two-loop integral topologies}
\label{fig:topology}
\end{figure}

 The renormalized massless form factors $\td{c}^{R;q,g}_{Ii}$ can be extracted according to 
\begin{align}\label{eq:formfactorRmassless}
 \td{c}^{R;q,g}_{Ii}=\sum_{j}\frac{\lt(\td{\bm{D}}_{q,g}^{-1}\rt)_{ij}}{\braket{c^{q,g}_I|c^{q,g}_I}}\lt[\bra{\td{d}^{q,g}_j}\otimes\braket{c^{q,g}_I|\mC{\td{M}}^{R}_{q,g}}\rt]\,,
\end{align}
where $\td{\bm{D}}^{q,g}_{ij} = \braket{\td{d}^{q,g}_i|\td{d}^{q,g}_j}$. We generate the massless amplitudes using \texttt{FeynArts} \cite{Hahn:2000kx}, and manipulate the expressions with \texttt{FeynCalc} \cite{Mertig:1990an, Shtabovenko:2016sxi, Shtabovenko:2020gxv} and \texttt{FORM} \cite{Vermaseren:2000nd}. After applying the projectors in Eq.~\eqref{eq:formfactorRmassless}, the one-loop and two-loop massless form factors can be expressed as linear combinations of one-loop and two-loop scalar Feynman integrals, respectively. The two-loop scalar integrals are categorized into 180 integral families. Taking into account permutations of external momenta, there are four independent integral topologies. These integrals have been considered in Refs.~\cite{Gehrmann:2015bfy,Papadopoulos:2015jft,Gehrmann:2018yef,Chicherin:2018mue,Abreu:2018rcw,Abreu:2018aqd,Chicherin:2018old,Badger:2019djh}. Following the notation of Ref.~\cite{Chicherin:2020oor}, we define the two-loop integral families $G_{\tau,\sigma}$ for each topology $\tau$ in permutation $\sigma$ as
\begin{equation}
G_{\tau,\sigma}[\vec{a}] \equiv e^{2\eps\gamma_E}\int\fc{d^Dl_1}{\text{i}\pi^{\fc{D}{2}}}\fc{d^Dl_2}{\text{i}\pi^{\fc{D}{2}}}\fc{1}{\vec{\bm{D}}^{\vec{a}}_{\tau,\sigma}} \,, \quad \vec{\bm{D}}_{\tau,\sigma}^{\vec{a}}=\prod_i D_{\tau,\sigma,i}^{a_i} \,.
\end{equation}
For each of the four independent integral topologies, we choose the standard permutation $\sigma_0 = (1,2,3,4,5)$, and define the sets $\vec{\bm{D}}_{\tau,\sigma_0}$ as
\begin{equation} \label{eq:denominators}
  \begin{tabular}{rllll}
  \hline
      &\(\vec{\bm{D}}_{a,\sigma_0} \) &\(\vec{\bm{D}}_{b,\sigma_0} \) &\(\vec{\bm{D}}_{c,\sigma_0} \) &\(\vec{\bm{D}}_{d,\sigma_0} \)\\
  \hline
   1 & \((l_1)^2\) & \((l_1)^2\) & \((l_1)^2\) & \((l_1)^2\) \\
   2 & \((l_1+p_1)^2\) & \((l_1-p_1)^2\) & \((l_1-p_1)^2\) & \((l_1-p_1)^2\) \\
   3 & \((l_1+p_1+p_2)^2\) & \((l_1-p_1-p_2)^2\) & \((l_1-p_1-p_2)^2\) & \((l_1-p_1-p_2)^2\) \\
   4 & \((l_1-p_4-p_5)^2\) & \((l_1+p_4+p_5)^2\) & \((l_2)^2\) & \((l_2)^2\) \\
   5 & \((l_2)^2\) & \((l_2)^2\) & \((l_2+p_4+p_5)^2\) & \((l_2+p_4+p_5)^2\)  \\
   6 & \((l_2-p_4-p_5)^2\) & \((l_2+p_5)^2\) & \((l_2+p_5)^2\) &\( (l_2-p_1-p_2)^2\)  \\
   7 & \((l_2-p_5)^2\) & \((l_1-l_2)^2\) & \((l_1-l_2)^2\) &  \((l_2+p_5)^2\)  \\
   8 & \((l_1-l_2)^2\) & \((l_1-l_2+p_4)^2\) & \((l_1-l_2+p_3)^2\) & \((l_1-l_2)^2\)  \\
 \hline
   9 & \((l_1-p_5)^2\) & \((l_2-p_1)^2\) & \((l_1+p_5)^2\) &\((l_1-l_2+p_3)^2\)  \\
   10 & \((l_2+p_1)^2\) & \((l_2-p_1-p_2)^2\) & \((l_2-p_1)^2\) & \((l_1+p_5)^2\) \\
   11 & \((l_2+p_1+p_2)^2\) & \((l_2+p_4+p_5)^2\) & \((l_2-p_1-p_2)^2\) & \((l_2-p_1)^2\) \\
  \end{tabular}
\end{equation}
The corresponding diagrams are depicted in Figure~\ref{fig:topology}.

We now need to perform IBP reduction for these four independent topologies. We note that the most difficult part is the reduction of the top-sector in topology-DP, where the total power of irreducible scalar products in the numerator can be up to $5$. The IBP relations in this top-sector have been given in the literature \cite{Bendle:2019csk, Boehm:2020ijp, Bendle:2021ueg}. For the sub-sectors in this topology, and the other three topologies, we perform IBP reduction using the program package \texttt{Kira} \cite{Maierhoefer:2017hyi, Klappert:2020nbg} with the help of \texttt{FireFly} \cite{Klappert:2019emp, Klappert:2020aqs}. We find that the top-sector of topology-HT is reducible and the masters in this topology can be obtained from topology-PB with a suitable permutation of external momenta. The size of these reduction relations is around 15~GB. To simplify subsequent calculations, we follow the suggestion of \cite{Bendle:2019csk}, and express the integrals in terms of the uniform transcendentality (UT) basis given in \cite{Chicherin:2020oor} for each topology.\footnote{Note that the UT bases for the top-sector of topology-DP given in \cite{Bendle:2019csk, Boehm:2020ijp, Bendle:2021ueg} are different from those in \cite{Chicherin:2020oor}. They are related by linear transformations.} The IBP coefficients are further simplified using the program package \texttt{MultivariateApart} \cite{Heller:2021qkz}. As a result, the size of the reduction relations becomes about 2.2~GB.
Finally, the UT bases can be evaluated using the program package \texttt{PentagonMI} \cite{Chicherin:2020oor}. With the same program, we also compute the one-loop UT bases up to weight 4, which are necessary for calculating the one-loop form factors up to order $\epsilon^2$.

\section{Squared amplitudes and mapping between massive and massless phase-space points}\label{sec:squaredamp}

We have noted in Eq.~\eqref{eq:QCDfac} that the factorization formula receives power corrections. As such, there are ambiguities of $\mathcal{O}(m^2)$ among different conventions of using the formula. In particular, when we apply the formula to compute the physical cross sections, there are two kinds of ambiguities that we'll discuss in this Section.

\subsection{Squared amplitudes}

The first kind of ambiguity comes from how to apply the formula to compute the squared amplitudes. Since the tree-level and one-loop massive amplitudes can be calculated exactly, the only ambiguity in the NNLO squared amplitudes lies in the interference between the two-loop amplitudes and the tree-level ones. We will pick two particular schemes and compare the outcome, while keeping in mind that other choices are possible.

The first scheme is that we strictly square the approximate amplitudes $\ket{\bar{\mC{M}}^R_{q,g}}$ in Eq.~\eqref{eq:QCDfac1}, and extract the NNLO terms, which are simply $2\operatorname{Re}\braket{\bar{\mC{M}}^{(0)R}_{q,g}|\bar{\mC{M}}^{(2)R}_{q,g}}$. Here, $\ket{\bar{\mC{M}}^{(0)R}_{q,g}}$ are the same as the massless tree-level amplitudes $\ket{\td{\mC{M}}^{(0)R}_{q,g}}$. The only mass information is contained in the factors of $\zcal_{[j]}^{(m|0)}$ and the soft function $\bm{\scal}$. We will refer to this scheme as the ``massless scheme''.

The second scheme is a bit more involved. We introduce a modified version of the approximate amplitudes:
\begin{multline}
\label{eq:QCDfac3}
\Ket{\hat{\mC{M}}^R_{q,g}\lt(\epsilon,\{p\},m_t,m_H,\mu\rt)}=\zcal_{[q,g]}^{(m|0)}(\epsilon,m_t,\mu) \, \zcal_{[t]}^{(m|0)}(\epsilon,m_t,\mu)
\\
\times \bm{\scal}(\epsilon,\{\tilde{p}\},m_t,\mu) \sum_{I,i}\td{c}^{R;q,g}_{Ii}\ket{c_I^{q,g}} \,\otimes \, \ket{\hat{d}_i^{q,g}} 
\end{multline}
where $\ket{\hat{d}_i^{q,g}}$ are the massive version of $\ket{\td{d}_i^{q,g}}$. That is, the spinors and the momenta of the (anti-)top quark and the Higgs boson are treated as massive. The difference between $\hat{\mC{M}}^R_{q,g}$ and $\bar{\mC{M}}^R_{q,g}$ is power-suppressed. We can then define the interference between the approximate two-loop amplitudes with the fully massive tree-level amplitudes via $\braket{\mC{M}^{(0)R}_{q,g}|\hat{\mC{M}}^{(2)R}_{q,g}}$. In this setup, only the two-loop form factors $\td{c}^{(2)R;q,g}_{Ii}$ are purely massless and are calculated according to Eq.~\eqref{eq:formfactorRmassless}. The other quantities, including the tree-level amplitudes $\mC{M}^{(0)R}_{q,g}$ and the spin (Lorentz) structures $\ket{\hat{d}_i^{q,g}}$, all contain mass information. Therefore, we will refer to this scheme as the ``semi-massive scheme''.

\subsection{Mapping between massive and massless phase-space points}

The second kind of ambiguity lies in the relation between the massive external momenta $p_i$ and the massless external momenta $\tilde{p}_i$ in Eq.~\eqref{eq:QCDfac1}. In practice, when performing phase-space integration, we first generate a set of massive momenta $p_i$. We then need to uniquely fix a corresponding set of $\tilde{p}_i$ that differ from $p_i$ only by power corrections. Our convention is simple: we set the directions of the 3-momenta unchanged between $p_i$ and $\tilde{p}_i$, and rescale the norms of the 3-momenta together with the energy parts to fulfill the massless on-shell condition while keeping momentum conservation.

To be more precise, since $p_1$ and $p_2$ are already massless, we will simply choose $\tilde{p}_1=p_1$ and $\tilde{p}_2=p_2$. Therefore we have $\td{s}_{12}=s_{12}$. We now parameterize $p_i$ in the center-of-mass frame as
\begin{align}\label{eq:momentamassive}
p_1&=\frac{\sqrt{s_{12}}}{2}\lt(1,0,0,1\rt)\,,\nn\\
p_2&=\frac{\sqrt{s_{12}}}{2}\lt(1,0,0,-1\rt)\,,\nn\\
p_3&=\lt(\sqrt{m_t^2+q_3^2},q_3\sin\theta_3\sin\phi_3,q_3\sin\theta_3\cos\phi_3,q_3\cos\theta_3\rt)\,,\nn\\
p_4&=\lt(\sqrt{m_t^2+q_4^2},q_4\sin\theta_4\sin\phi_4,q_4\sin\theta_4\cos\phi_4,q_4\cos\theta_4\rt)\,,\nn\\
p_5&=\lt(\sqrt{m_H^2+q_5^2},q_5\sin\theta_5,0,q_5\cos\theta_5\rt) \,,
\end{align}
where $q_i$ is the norm of the spatial components of $p_i$, $\theta_i$ is the polar angle and $\phi_i$ is the azimuthal angle. For convenience and without loss of generality, we have set $\phi_5=0$.
Due to momentum conservation and the on-shell conditions, there are 7 independent parameters in the right-hand side of Eq.~\eqref{eq:momentamassive}, which are chosen as $s_{12}$, $m_t$, $m_H$, $q_5$, $\theta_3$, $\phi_3$ and $\theta_5$. Their values satisfy the physical constraints
\begin{align}
s_{12} \ge (2m_t&+m_H)^2\,, \quad 0\le q_5\le q_{5,\rm max}=\frac{\sqrt{s_{12}-(2m_t+m_H)^2}\sqrt{s_{12}-(2m_t-m_H)^2}}{2\sqrt{s_{12}}}\,,\nn \\
&m_t >0 \,, \quad m_H>0 \,, \quad 0\le \theta_3 \le \pi\,, \quad 0\le \phi_3 < 2\pi\,, \quad 0\le \theta_5 \le \pi \,.
\end{align}

For the massless momenta $\td{p}_i$, we can similarly parameterize them as
\begin{align}
\td p_1&=\frac{\sqrt{\td s_{12}}}{2}\lt(1,0,0,1\rt)\,,\nn\\
\td p_2&=\frac{\sqrt{\td s_{12}}}{2}\lt(1,0,0,-1\rt)\,,\nn\\
\td p_3&=\lt(\td q_3,\td q_3\sin\td \theta_3\sin\td \phi_3,\td q_3\sin\td \theta_3\cos\td \phi_3,\td q_3\cos\td \theta_3\rt),\nn\\
\td p_4&=\lt(\td q_4,\td q_4\sin\td \theta_4\sin\td \phi_4,\td q_4\sin\td \theta_4\cos\td \phi_4,\td q_4\cos\td \theta_4\rt),\nn\\
\td p_5&=\lt(\td q_5,\td q_5\sin\td \theta_5,0,\td q_5\cos\td \theta_5\rt) \,.
\end{align}
As mentioned before, we will set $\td s_{12}=s_{12}$, $\td \theta_i = \theta_i$ and $\td \phi_i = \phi_i$. The remaining parameters $\td q_i$ can then be solved from momentum conservation. They can be written as
\begin{align}
\td q_3&=\fc{\sqrt{s_{12}}\cos\phi_4\sin\theta_5\sin\theta_4}{\cos\phi_4\sin\theta_4(\sin\theta_5-\sin\theta_3\sin\phi_3)-\cos\phi_3\sin\theta_3(\sin\theta_5-\sin\theta_4\sin\phi_4)}\,,\nn\\
\td q_4&=\fc{-\sqrt{s_{12}}\cos\phi_3\sin\theta_5\sin\theta_3}{\cos\phi_4\sin\theta_4(\sin\theta_5-\sin\theta_3\sin\phi_3)-\cos\phi_3\sin\theta_3(\sin\theta_5-\sin\theta_4\sin\phi_4)}\,,\nn\\
\td q_5&=\fc{-\sqrt{s_{12}}\sin\theta_3\sin\theta_4\sin(\phi_3-\phi_4)}{\cos\phi_4\sin\theta_4(\sin\theta_5-\sin\theta_3\sin\phi_3)-\cos\phi_3\sin\theta_3(\sin\theta_5-\sin\theta_4\sin\phi_4)}\,.\nn
\end{align}

\section{Numeric results}\label{sec:numeric}

We are now ready to present the numeric results for the two-loop amplitudes based on our approximate formula. Before that, we briefly discuss the choice of the renormalization scale $\mu$. The kinematic configuration that we're considering involves vastly different scales $|s_{ij}| \gg m_t^2$. In this case, there will be large logarithms such as $\ln(s_{ij}/m_t^2)$ no matter what we choose as $\mu$. The standard way to deal with this problem is to resum these logarithms using renormalization group equations, based on the factorization formula \eqref{eq:QCDfac} (see, e.g., \cite{Mele:1990cw, Ferroglia:2012ku, Ferroglia:2013awa, Pecjak:2016nee, Czakon:2018nun}). However, in this work we are only using the factorization formula to provide an approximation of the two-loop amplitudes. Therefore, we are free to use any scale $\mu$ to present our results. We will by default set $\mu = m_t$ in the following, and will compare with other scale choices. For the other input parameters, we set $m_t = \SI{173}{\GeV}$ and $m_H = \SI{125}{\GeV}$.

In order to assess the validity of the approximate formula, we will first compare the approximate results with the exact ones. These exact results include the NLO squared amplitudes up to the finite parts, and the IR poles of the NNLO squared amplitudes \cite{Chen:2022nxt}. The NLO squared amplitudes come from the interference between tree-level and one-loop amplitudes, which can be obtained from the literature as well as various program packages \cite{Beenakker:2001rj, Beenakker:2002nc, Reina:2001sf, Reina:2001bc, Dawson:2002tg, Genovese:2003dp, Alwall:2014hca, Frederix:2018nkq, Cullen:2011ac, GoSam:2014iqq, Buccioni:2019sur, Cascioli:2011va, Buccioni:2017yxi, Actis:2016mpe, Denner:2017wsf}. Since we want to compare at the level of different color structures, we have recomputed the exact NLO results. The NNLO squared amplitudes receive two contributions: the one-loop-squared amplitudes, and the interference between tree-level and two-loop amplitudes. The finite part of the latter contribution is the genuinely new result of this work. Hence, for convenience, we will drop the one-loop-squared amplitudes and only compare the interference terms.

\begin{figure}[t!]
\centering  
\subfigure{
\label{fig:qqnloeps0}
\includegraphics[width=0.48\textwidth]{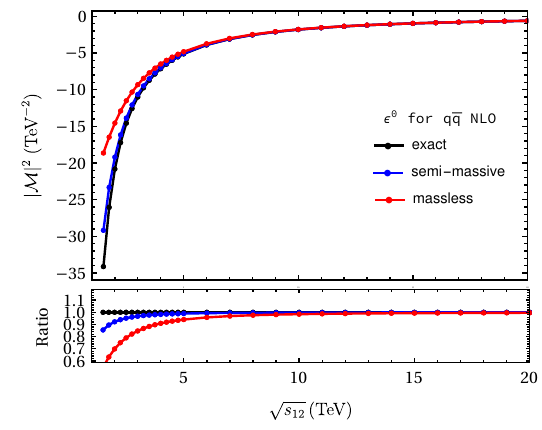}}
\subfigure{
\label{fig:ggnloeps0}
\includegraphics[width=0.48\textwidth]{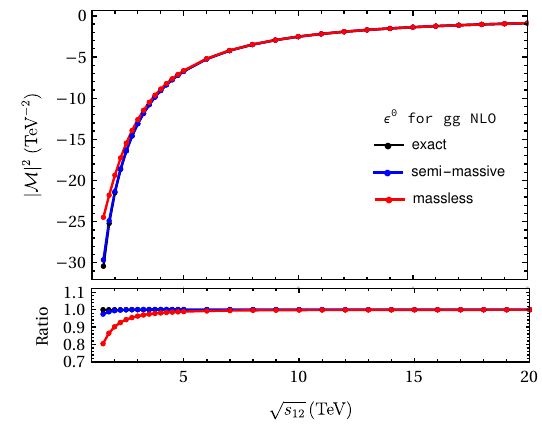}} \;
\subfigure{
\label{fig:qqnnloepsm1}
\includegraphics[width=0.48\textwidth]{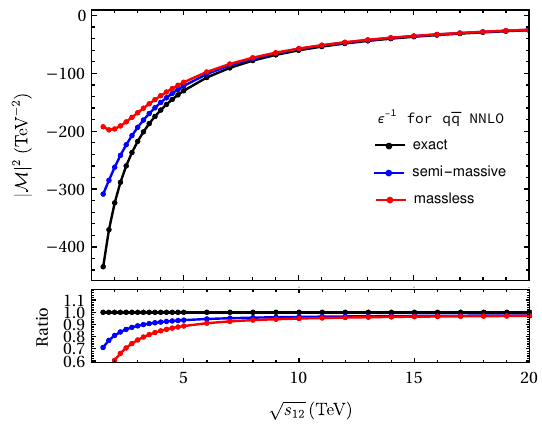}}
\subfigure{
\label{fig:ggnnloepsm1}
\includegraphics[width=0.48\textwidth]{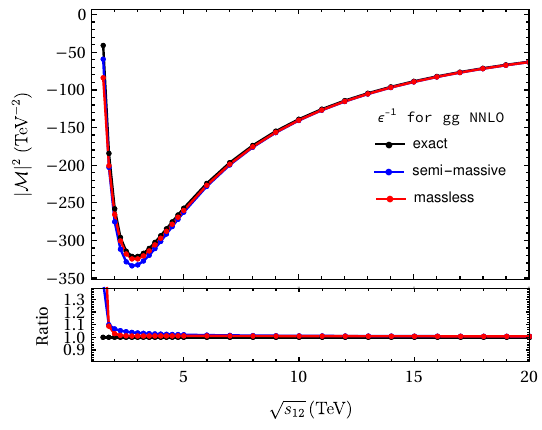}}
\caption{The squared amplitudes at $\mC{O}(\epsilon^0)$ for NLO and $\mC{O}(\epsilon^{-1})$ for NNLO in both the $q\bar{q}$ and $gg$ channels, as a function of the partonic center-of-mass energy $\sqrt{s_{12}}$. The other kinematic variables are chosen as $\theta_3 = 14\pi/29$, $\phi_3=34\pi/29$, $\theta_5=15\pi/29$ and $q_5=20\,q_{5,\rm max}/29$. We have included the spin- and color-average factors of $1/36$ for the $q\bar{q}$ channel and $1/256$ for the $gg$ channel. The lower panel in each plot shows the ratios between the approximate results and the exact ones.}
\label{fig:diffscheme}
\end{figure}

We first show in Fig.~\ref{fig:diffscheme} the numeric results at $\mC{O}(\eps^0)$ for NLO (upper plots) and $\mC{O}(\eps^{-1})$ for NNLO (lower plots) squared amplitudes for both $q\bar{q}$ (left plots) and $gg$ (right plots) channels. We change the partonic center-of-mass energy $\sqrt{s_{12}}$ from about $\SI{1.5}{\TeV}$ to $\SI{20}{\TeV}$, and fix the other kinematic variables as $\theta_3 = 14\pi/29$, $\phi_3=34\pi/29$, $\theta_5=15\pi/29$ and $q_5=20\,q_{5,\rm max}/29$. The black lines are exact results, while the red and blue curves are approximate results using the two different schemes discussed in Sec.~\ref{sec:squaredamp}. As expected, the approximation works better towards higher energies both at NLO and NNLO. We observe that the semi-massive scheme (the blue curves) generally has a better performance than the massless scheme. Such an effect can be attributed to the fact that the semi-massive scheme incorporates some of the power-suppressed contributions. However, this should be taken with a grain of salt since these beyond-LP terms can only be reliably obtained by a real calculation at the corresponding orders. Overall, the semi-massive scheme provides a reasonable approximation to the exact results down to $\sqrt{s_{12}} \sim \SI{2}{\TeV}$ in all cases.

Interestingly, we observed from Fig.~\ref{fig:diffscheme} that the approximation works better in the $gg$ channel than in the $q\bar{q}$ one. In particular, the relative difference in the $q\bar{q}$ channel between the result in the massless scheme and the exact one is much bigger than the naive estimation of the order $(m_t^2,m_H^2)/|s_{ij}|$. In fact, this effect is already present at LO, and is inherited to higher orders. The tree-level squared amplitude in the $q\bar{q}$ channel is a rational function of $m_t$, $m_H$ and $s_{ij}$. When expanded in the small-mass limit, there is an accidental cancellation at the LP. Numerically, the LP contribution is similar to the next-to-leading power (NLP) one. These power corrections at LO are dropped in the massless scheme, but are kept in the semi-massive scheme. This is the reason why the semi-massive scheme works much better than the massless scheme in the $q\bar{q}$ channel.

\begin{figure}[t!]
\centering  
\subfigure{
\label{fig:qqnloeps0t5}
\includegraphics[width=0.48\textwidth]{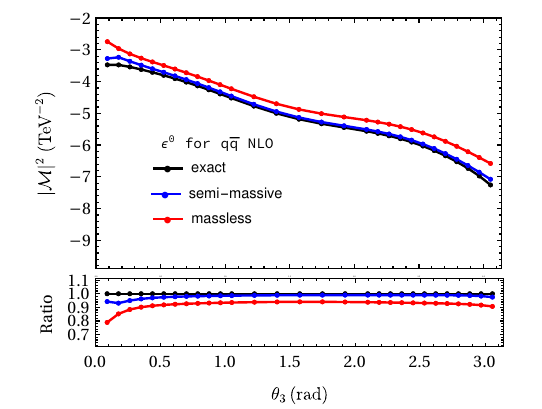}}
\subfigure{
\label{fig:ggnloeps0t5}
\includegraphics[width=0.48\textwidth]{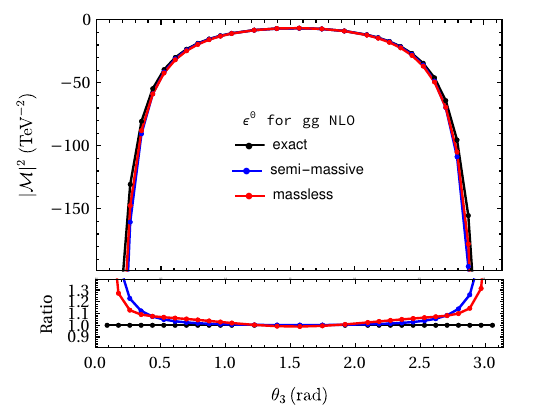}} \;
\subfigure{
\label{fig:qqnnloepsm1t5}
\includegraphics[width=0.48\textwidth]{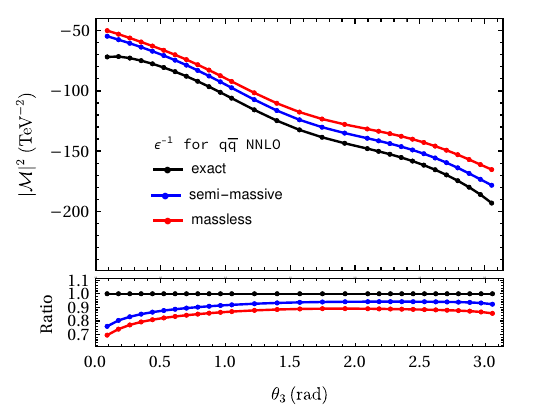}}
\subfigure{
\label{fig:ggnnloepsm1t5}
\includegraphics[width=0.48\textwidth]{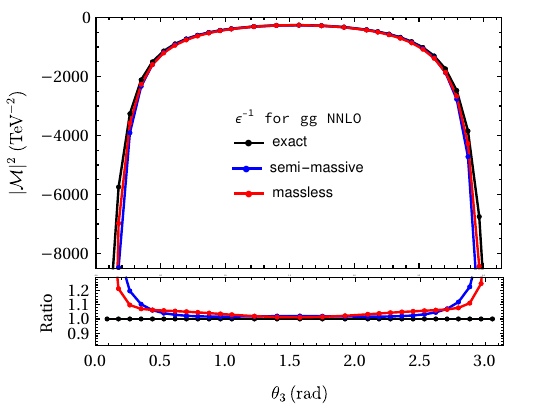}}
\caption{The squared amplitudes at $\mC{O}(\epsilon^0)$ for NLO and $\mC{O}(\epsilon^{-1})$ for NNLO in both the $q\bar{q}$ and $gg$ channels, as a function of the angle parameter $\theta_3$. The other kinematic variables are chosen as $\sqrt{s_{12}}=\SI{5}{\TeV}$, $\phi_3=34\pi/29$, $\theta_5=15\pi/29$ and $q_5=20\,q_{5,\rm max}/29$.}
\label{fig:diffschemetheta3s5}
\end{figure} 

We now turn to investigate the behaviors of the approximate results when varying the angle $\theta_3$ between $\vec{p}_3$ (of the top quark) and $\vec{p}_1$. The other parameters are fixed to be $\sqrt{s_{12}}= \SI{5}{\TeV}$, $\phi_3=34\pi/29$, $\theta_5=15\pi/29$ and $q_5=20\,q_{5,\rm max}/29$. It can be expected that when $\theta_3 \to 0$ or $\pi$, the approximation will become worse due to the smallness of either $|s_{13}|$ or $|s_{23}|$. Indeed, as is evident from Fig.~\ref{fig:diffschemetheta3s5}, the approximate results deviate from the exact ones at the two ends of the spectrum. Especially in the $gg$ channel, the tree-level amplitudes contain $1/(s_{13}-m_t^2)$ and $1/(s_{23}-m_t^2)$ propagators. Whether or not one takes $m_t \to 0$ will make a huge difference for these terms. This explains the wild behaviors of the blue and red curves for the $gg$ channel at the left and right ends. Nevertheless, the semi-massive scheme still provides a reasonable approximation in a large portion of the phase space in both the $q\bar{q}$ and $gg$ cases.

\begin{figure}[t!]
\centering  
\subfigure{
\label{fig:qqnloeps0thetah5}
\includegraphics[width=0.48\textwidth]{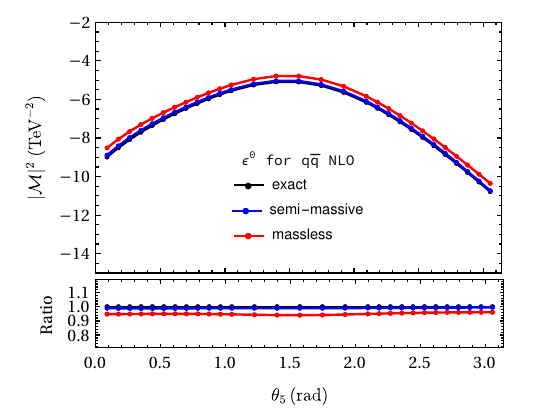}}
\subfigure{
\label{fig:ggnloeps0thetah5}
\includegraphics[width=0.48\textwidth]{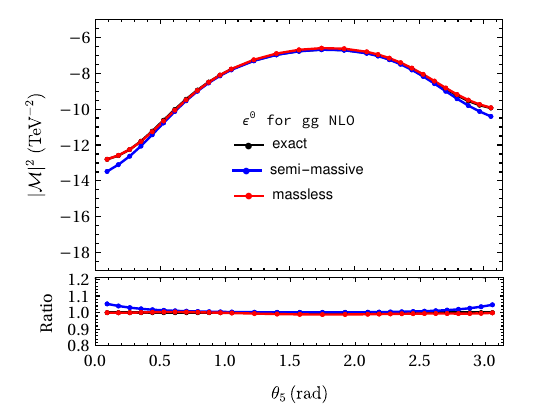}} \;
\subfigure{
\label{fig:qqnnloepsm1thetah5}
\includegraphics[width=0.48\textwidth]{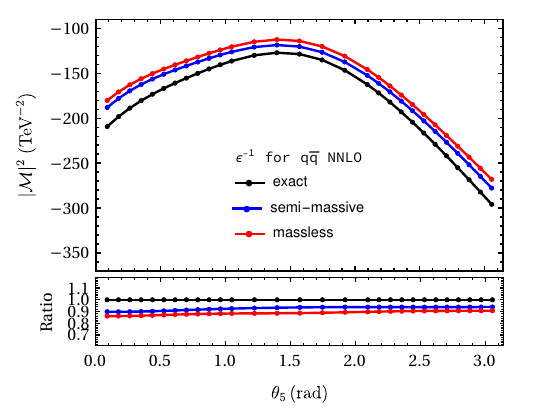}}
\subfigure{
\label{fig:ggnnloepsm1thetah5}
\includegraphics[width=0.48\textwidth]{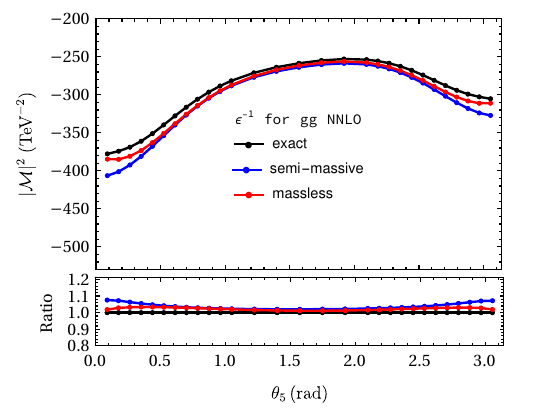}}
\caption{The squared amplitudes as a function of the angle parameter $\theta_5$. The other kinematic variables are chosen as: $\sqrt{s_{12}}=\SI{5}{\TeV}$, $\theta_3=14\pi/29$, $\phi_3=34\pi/29$ and $q_5=20\,q_{5,\rm max}/29$.}
\label{fig:diffschemethetah5s5}
\end{figure}

The situation is slightly different in the case of $\theta_5$, the angle between $\vec{p}_5$ (of the Higgs boson) and $\vec{p}_1$. Since the Higgs boson is color neutral, it is expected that the QCD corrections should have a mild dependence on its momentum. This can be seen from Fig.~\ref{fig:diffschemethetah5s5}. The approximation is rather good in the whole range of $\theta_5$, as long as $\theta_3$ is in the wide-angle region. From this and the previous plots, we find that the semi-massive scheme has an overall better performance than the massless scheme. Therefore, we will use the semi-massive scheme as the default to present our numerical results in the following.

\begin{figure}[t!]
\centering  
\subfigure{
\includegraphics[width=0.48\textwidth]{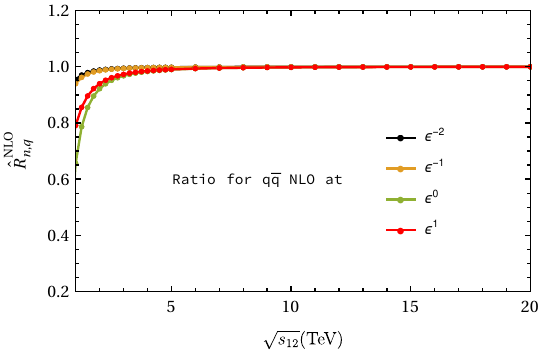}}
\subfigure{
\includegraphics[width=0.48\textwidth]{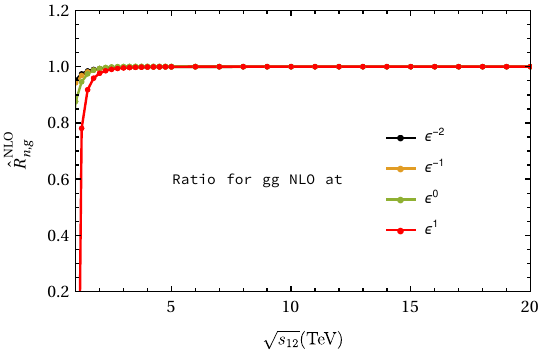}}
\subfigure{
\includegraphics[width=0.48\textwidth]{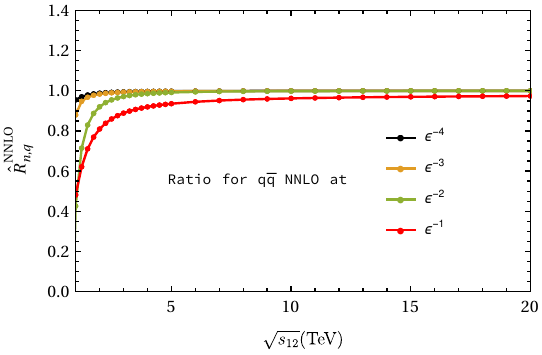}}
\subfigure{
\includegraphics[width=0.48\textwidth]{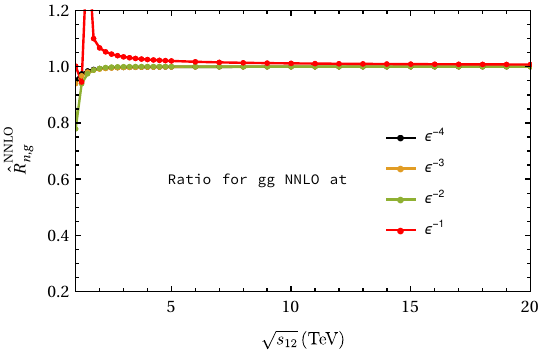}}
\caption{The ratios between the approximate results in the semi-massive scheme and the exact ones at different orders in $\epsilon$, as functions of $\sqrt{s_{12}}$. The phase-space parameters are the same as those in the Fig.~\ref{fig:diffscheme}.}
\label{fig:rationloshat}
\end{figure}

\begin{figure}[t!]
\centering  
\subfigure{
\includegraphics[width=0.48\textwidth]{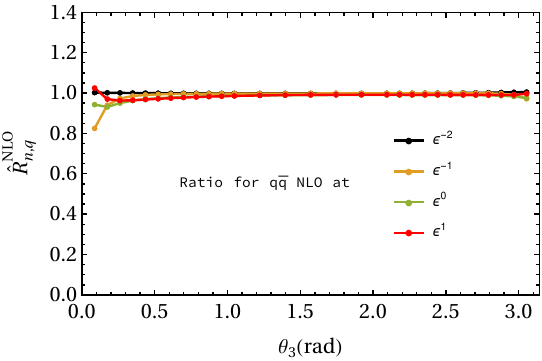}}
\subfigure{
\includegraphics[width=0.48\textwidth]{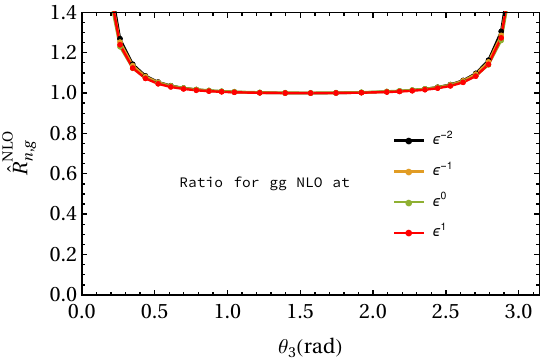}}
\subfigure{
\includegraphics[width=0.48\textwidth]{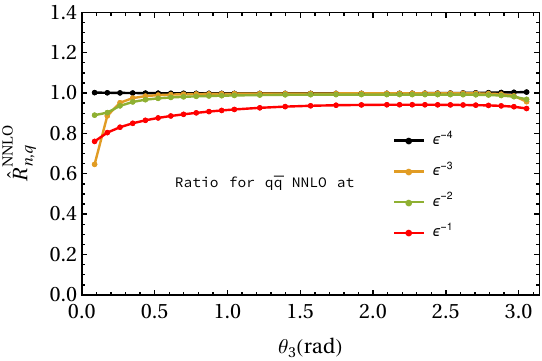}}
\subfigure{
\includegraphics[width=0.48\textwidth]{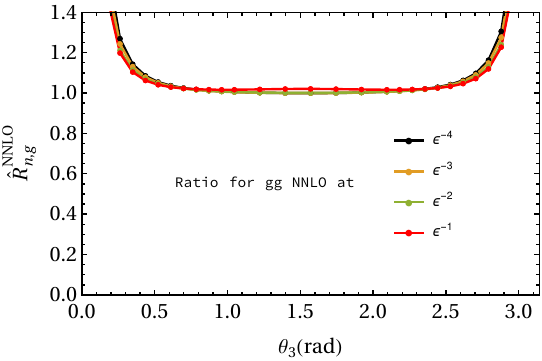}}
\caption{The ratios between the approximate results in the semi-massive scheme and the exact ones at different orders in $\epsilon$, as functions of $\theta_3$. The other phase-space parameters are the same as those in the Fig.~\ref{fig:diffschemetheta3s5}.}
\label{fig:rationlotheta5}
\end{figure}

Now we turn to consider the NLO and NNLO squared amplitudes at different orders in $\epsilon$ in the semi-massive scheme. We define the following ratios between the approximate results and the exact ones at each order in $\epsilon$:
\begin{align}
\hat{R}_{n,q/g}^{\text{NLO}}=\frac{\mathrm{Re}\braket{\mC{M}^{(0)R}_{q/g}|\hat{\mC{M}}^{(1)R}_{q/g}}\bigg|_{\epsilon^n}}{\mathrm{Re}\braket{\mC{M}^{(0)R}_{q/g}|\mC{M}^{(1)R}_{q/g}}\bigg|_{\epsilon^n}}\,, \quad \hat{R}_{n,q/g}^{\text{NNLO}}=\frac{\mathrm{Re}\braket{\mC{M}^{(0)R}_{q/g}|\hat{\mC{M}}^{(2)R}_{q/g}}\bigg|_{\epsilon^n}}{\mathrm{Re}\braket{\mC{M}^{(0)R}_{q/g}|\mC{M}^{(2)R}_{q/g}}\bigg|_{\epsilon^n}}\,,
\end{align}
where $\ket{\hat{\mC{M}}^{(i)R}_{q/g}}$ are defined in Eq.~\eqref{eq:QCDfac3} and $\ket{\mC{M}^{(i)R}_{q/g}}$ are the exact massive amplitudes. Note that $\ket{\mC{M}^{(1)R}_{q/g}}$ are known up to $\epsilon^1$ and $\ket{\mC{M}^{(2)R}_{q/g}}$ are known up to $\epsilon^{-1}$ in \cite{Chen:2022nxt}. 
We show $\hat{R}_{n,q/g}^{\text{NLO}}$ and $\hat{R}_{n,q/g}^{\text{NNLO}}$ as functions of $\sqrt{s_{12}}$ in Fig.~\ref{fig:rationloshat} with the same choices of phase-space parameters as in Fig.~\ref{fig:diffscheme}. It can be seen that, in general, the approximation works better for the coefficients of higher poles than those of lower ones. This can be expected since the higher poles are generically related to lower order (in $\alpha_s$) amplitudes and anomalous dimensions.
In Fig.~\ref{fig:rationlotheta5}, we show $\hat{R}_{n,q,g}^{\text{NLO}}$ and $\hat{R}_{n,q,g}^{\text{NNLO}}$ as functions of $\theta_3$, with the same phase-space points as Fig.~\ref{fig:diffschemetheta3s5}. We observe that the approximation is rather good for most values of $\theta_3$ in the $q\bar{q}$ region except when $\theta_3 \to 0$, where the $1/\epsilon$ term at NLO and the $1/\epsilon^3$ term at NNLO behave weirdly. This behavior is partly due to the incomplete information of the power-suppressed contributions in the semi-massive scheme, which already appears at the tree-level. In the $gg$ channel, the approximation works perfectly in the central region, while breaking down in the forward regions $\theta_3 \to 0$ or $\pi$. This can be expected since in these regions the propagator denominator $s_{13} - m_t^2$ or $s_{23} - m_t^2$ cannot be described reliably using small-mass expansion.

\begin{figure}[t!]
\centering  
\subfigure{
\includegraphics[width=0.48\textwidth]{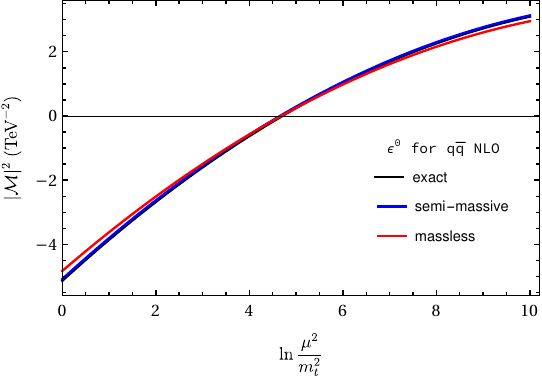}}
\subfigure{
\includegraphics[width=0.48\textwidth]{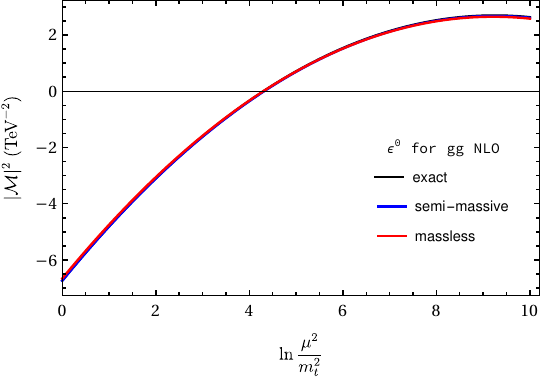}}
\subfigure{
\includegraphics[width=0.48\textwidth]{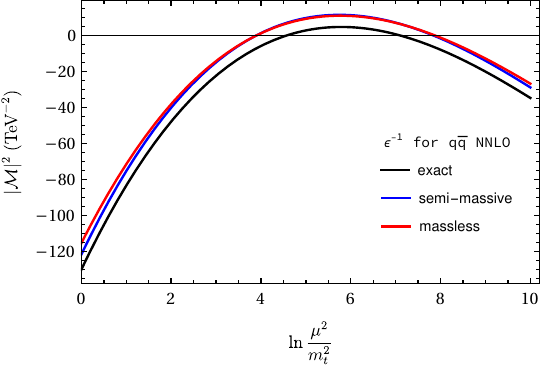}}
\subfigure{
\includegraphics[width=0.483\textwidth]{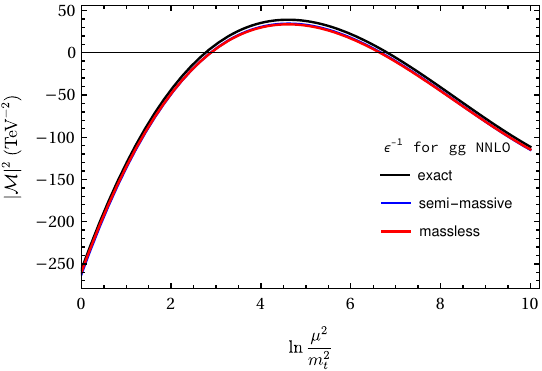}}
\caption{The squared amplitudes as a function of the renormalization scale $\mu$. Here we show the $\epsilon^0$ coefficients at NLO and the $\epsilon^{-1}$ coefficients at NNLO. The center-of-mass energy is chosen as $\sqrt{s_{12}}=\SI{5}{\TeV}$ and other phase-space parameters are the same as those in Fig.~\ref{fig:diffscheme}.}
\label{fig:squaredamplogmu5TeV}
\end{figure}

The above results are all computed using the default scale choice $\mu = m_t$. In Fig.~\ref{fig:squaredamplogmu5TeV}, we show the scale dependence of the squared amplitudes at NLO and NNLO. We can see that the quality of the approximation is nearly the same for different choices of $\mu$. It is worth noting that without the $n_h^1 n_l^0$ terms in our factorization formula computed in \cite{Wang:2023qbf}, the approximate results would deviate for $\mu \neq m_t$. This shows that our approximate formula has correctly captured the scale-dependence of the amplitudes.

\begin{figure}[t!]
\centering  
\subfigure{
\includegraphics[width=0.48\textwidth]{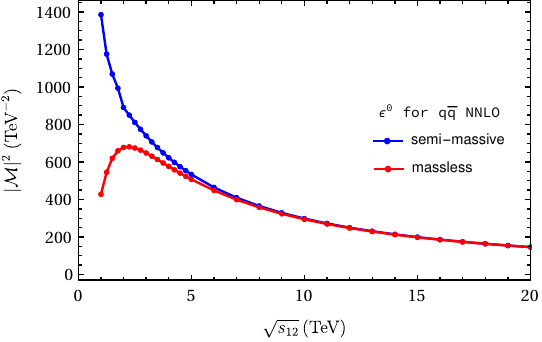}}
\subfigure{
\includegraphics[width=0.48\textwidth]{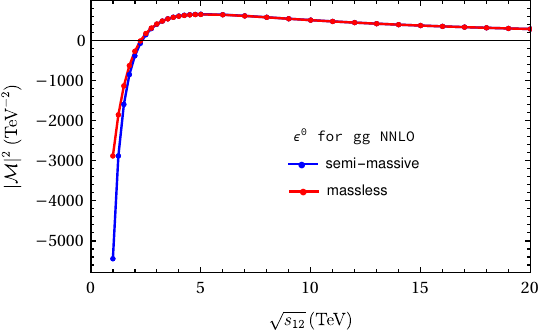}}
\subfigure{
\includegraphics[width=0.48\textwidth]{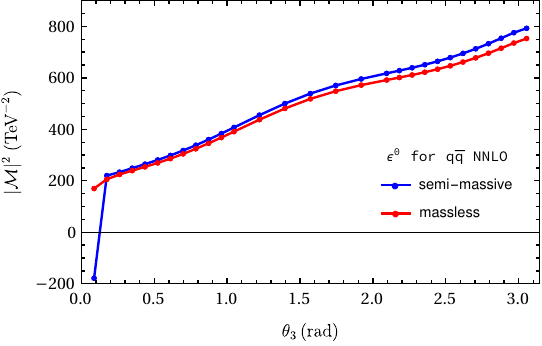}}
\subfigure{
\includegraphics[width=0.48\textwidth]{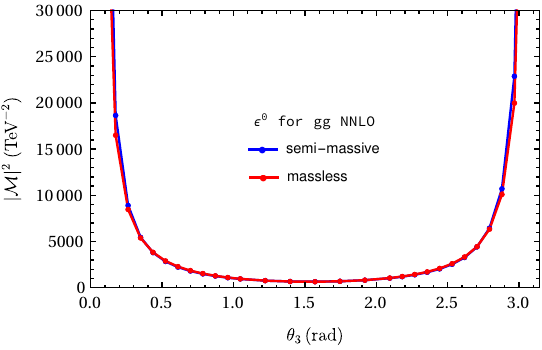}}
\subfigure{
\includegraphics[width=0.48\textwidth]{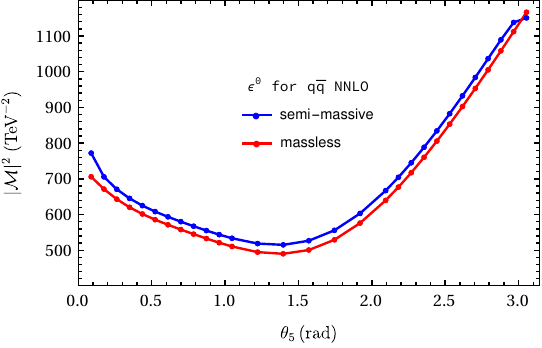}}
\subfigure{
\includegraphics[width=0.48\textwidth]{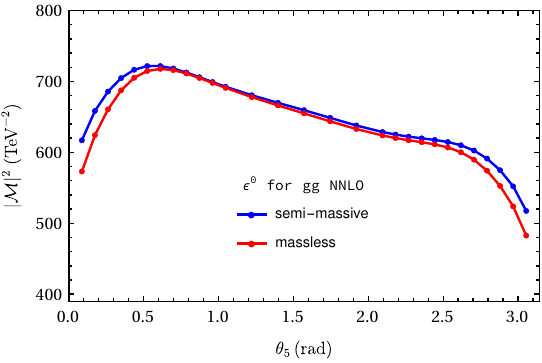}}
\caption{The finite parts of the NNLO squared amplitudes in the $q\bar{q}$ and $gg$ channels. The phase-space points follow those in Fig.~\ref{fig:diffscheme}, \ref{fig:diffschemetheta3s5} and \ref{fig:diffschemethetah5s5}.}
\label{fig:nnloeps0}
\end{figure}

We now come to the main results of this paper, namely the approximate predictions for the finite parts of the two-loop amplitudes for $t\bar{t}H$ production. We show in Fig.~\ref{fig:nnloeps0} the results as functions of $\sqrt{s_{12}}$, $\theta_3$ and $\theta_5$, respectively. The phase-space points are chosen following Fig.~\ref{fig:diffscheme}, \ref{fig:diffschemetheta3s5} and \ref{fig:diffschemethetah5s5}. Since the exact results of these finite terms are unknown, we present approximate results from both semi-massive and massless schemes for comparison. We find that the two results are quite similar at high energies, and start to deviate as we lower $\sqrt{s_{12}}$. Based on our findings in Fig.~\ref{fig:diffscheme}, we believe that the semi-massive scheme provides a more reliable approximation at low energies. The only exception is the $\theta_3 \to 0$ region, where the semi-massive result has a weird behavior as observed earlier in Fig.~\ref{fig:rationlotheta5}. In this region, the massless result seems to be more stable.

Besides the full squared amplitude as a whole, it is also interesting to study the squared amplitude decomposed by color coefficients. The color structure of $t\bar{t}H$ production is the same as that of $t\bar{t}$ production. Therefore, we employ the color decomposition according to \cite{Czakon:2007ej, Czakon:2007wk, Ferroglia:2009ii}:
\begin{align}
\label{eq:color_dec}
   2\,{\rm Re} \Braket{{\cal M}_{q}^{(0)}|{\cal M}_{q}^{(2)}}
   &= 2(N^2-1)\,\bigg( N^2 A^q + B^q + \frac{1}{N^2}\,C^q 
    + N n_l\,D_l^q + N n_h\,D_h^q \nonumber
    \\
   &+ \frac{n_l}{N}\,E_l^q 
    + \frac{n_h}{N}\,E_h^q + n_l^2 F_l^q + n_l n_h\,F_{lh}^q 
    + n_h^2 F_h^q \bigg) \,, \nonumber
    \\
   2\,{\rm Re} \Braket{{\cal M}_{g}^{(0)}|{\cal M}_{g}^{(2)}}
   &= (N^2-1)\,\bigg( N^3 A^g + N\,B^g + \frac{1}{N}\,C^g 
    + \frac{1}{N^3}\,D^g \nonumber \\
   &+ N^2 n_l\,E_l^g + N^2 n_h\,E_h^g 
    + n_l\,F_l^g + n_h\,F_h^g + \frac{n_l}{N^2}\,G_l^g 
    + \frac{n_h}{N^2}\,G_h^g \nonumber\\
   &+ N n_l^2 H_l^g + N n_l n_h\,H_{lh}^g 
    + N n_h^2 H_h^g + \frac{n_l^2}{N}\,I_l^g 
    + \frac{n_l n_h}{N}\,I_{lh}^g + \frac{n_h^2}{N}\,I_h^g \bigg) \,.
\end{align}
In Tables~\ref{tab:qqnum1}, \ref{tab:ggnum1}, \ref{tab:qqnum2}, \ref{tab:ggnum2}, \ref{tab:qqnum3} and \ref{tab:ggnum3}, we list the numeric values of the color-decomposed two-loop squared amplitudes at different orders in $\epsilon$. We show results at three representative phase-space points, characterized by the center-of-mass energy: relatively low, intermediate and very high. These data may serve as a cross-check for an exact evaluation of the two-loop squared amplitudes in the future.

\section{Conclusion}\label{sec:conclusion}

In this paper, we calculate the two-loop amplitudes for $t\bar{t}H$ production at hadron colliders in the high-energy boosted limit, where the scalar products of external momenta are much larger than the top-quark mass. We employ the factorization formula recently obtained in \cite{Wang:2023qbf}, that incorporates all two-loop contributions including those proportional to the number of heavy flavors $n_h$. Using this factorization formula, we write the massive amplitudes as a product of the corresponding massless ones with universal factors capturing collinear and soft dynamics. We compute the massless amplitudes by IBP reduction and the existing results for the master integrals. The massive amplitudes are then computed in two schemes: the massless scheme where all the external spinors are treated as massless, and the semi-massive scheme where the external spinors are taken to be massive. For the IR poles, we compare the approximate results in the two schemes with the exact results of \cite{Chen:2022nxt}. We find that the semi-massive scheme provides a better approximation in most of the phase-space regions. We also show that our approximation correctly reproduces the scale-dependence of the exact results. We then give our predictions for the finite parts of the two-loop squared amplitudes, which are the main new results of this work.

By combining the contributions from real emissions (including the double-real and real-virtual diagrams), our results can be utilized to compute the NNLO differential cross sections for this important process in the high-energy boosted limit.
One may further incorporate the low-energy approximations such as the threshold approximation and the soft-Higgs approximation, and interpolates into the intermediate regions. This can lead to a reasonable approximation across the whole phase space, and give reliable predictions for differential cross sections from low to high energies.

\acknowledgments
   This work was supported in part by the National Natural Science Foundation of China under Grant No. 12375097, 11975030 and 12147103, and the Fundamental Research Funds for the Central Universities.

\clearpage

\begin{table}[t!]
\begin{center}
\begin{tabular}{|l|r|r|r|r|r|}
    \hline
   \text{exact}&\multirow{2}*{$\eps^{-4}$}& \multirow{2}*{$\eps^{-3}$}&\multirow{2}*{$\eps^{-2}$}&\multirow{2}*{$\eps^{-1}$}&\multirow{2}*{$\eps^{0}$}\\ \cline{1-1}
   \text{semi-}&  &  &  &  & \\ \hline
   
\multirow{2}*{$A^q$}  &  $0.3557588$  &  $-2.232234$  &  $18.09587$  &  $-103.9532$  &    \\ \cline{2-6}  &  $0.3515265$  &  $-2.195621$  &  $16.85573$  &  $-96.44254$  &  $348.5042$\\ \hline
\multirow{2}*{$B^q$}  &  $-0.7115176$  &  $7.063085$  &  $-39.95150$  &  $64.86504$  &    \\ \cline{2-6}  &  $-0.7030530$  &  $6.967777$  &  $-37.89494$  &  $130.9224$  &  $-151.0649$\\ \hline
\multirow{2}*{$C^q$}  &  $0.3557588$  &  $-4.830851$  &  $9.261042$  &  $84.82862$  &    \\ \cline{2-6}  &  $0.3515265$  &  $-4.772156$  &  $8.634749$  &  $83.19236$  &  $-582.0071$\\ \hline
\multirow{2}*{$D^q_l$}  &      &  $-0.3557588$  &  $0.4713803$  &  $6.932929$  &    \\ \cline{2-6}  &      &  $-0.3515265$  &  $0.4624203$  &  $6.840171$  &  $-43.36795$\\ \hline
\multirow{2}*{$D^q_h$}  &      &      &  $-1.217990$  &  $8.594097$  &    \\ \cline{2-6}  &      &      &  $-1.513230$  &  $11.12860$  &  $-48.44219$\\ \hline
\multirow{2}*{$E^q_l$}  &      &  $0.3557588$  &  $-0.6853616$  &  $1.779527$  &    \\ \cline{2-6}  &      &  $0.3515265$  &  $-0.6768002$  &  $1.754250$  &  $-64.57500$\\ \hline
\multirow{2}*{$E^q_h$}  &      &      &  $1.217990$  &  $-0.5594025$  &    \\ \cline{2-6}  &      &      &  $1.513230$  &  $-3.261069$  &  $-57.86518$\\ \hline
\multirow{2}*{$F^q_l$}  &      &      &      &      &    \\ \cline{2-6}  &      &      &      &      &  $0.1730958$\\ \hline
\multirow{2}*{$F^q_{lh}$}  &      &      &      &      &    \\ \cline{2-6}  &      &      &      &      &  $0.3461917$\\ \hline
\multirow{2}*{$F^q_h$}  &      &      &      &      &    \\ \cline{2-6}  &      &      &      &      &  $0.1730958$\\ \hline
\multirow{2}*{Total}  &  $1.124373$  &  $-8.136548$  &  $56.27591$  &  $-323.8812$  &    \\ \cline{2-6}  &  $1.110997$  &  $-8.004476$  &  $51.79516$  &  $-262.2186$  &  $890.7902$\\ \hline
Ratio  &  $0.9881$  &  $0.9838$  &  $0.9204$  &  $0.8096$  &    \\ \hline

\end{tabular}
\caption{\label{tab:qqnum1} Color-decomposed NNLO squared amplitude in the $q\bar{q}$ channel at the phase-space point $(\sqrt{s_{12}}, \sqrt{|s_{13}|}, \sqrt{|s_{14}|}, \sqrt{|s_{23}|}, \sqrt{|s_{24}|}) = (2.0000, 1.1041, 1.1373, 1.1648, 1.1402)$~TeV, $m_t=\SI{173}{GeV}$, and $m_H=\SI{125}{GeV}$.}
\end{center}
\end{table}

\begin{table}[t!]
\begin{center}
\begin{tabular}{|l|r|r|r|r|r|}
    \hline
   \text{exact}&\multirow{2}*{$\eps^{-4}$}& \multirow{2}*{$\eps^{-3}$}&\multirow{2}*{$\eps^{-2}$}&\multirow{2}*{$\eps^{-1}$}&\multirow{2}*{$\eps^{0}$}\\ \cline{1-1}
   \text{semi-}&  &  &  &  & \\ \hline   

\multirow{2}*{$A^g$}  &  $6.189970$  &  $-38.61149$  &  $185.2507$  &  $-502.3539$  &    \\ \cline{2-6}  &  $6.140920$  &  $-38.20303$  &  $180.8479$  &  $-466.0329$  &  $277.4175$\\ \hline
\multirow{2}*{$B^g$}  &  $-9.662043$  &  $67.19163$  &  $-504.2190$  &  $2664.419$  &    \\ \cline{2-6}  &  $-9.528888$  &  $66.04951$  &  $-485.4858$  &  $2302.867$  &  $-8009.039$\\ \hline
\multirow{2}*{$C^g$}  &      &  $-26.39205$  &  $137.5097$  &  $122.6492$  &    \\ \cline{2-6}  &      &  $-26.08095$  &  $129.1071$  &  $-282.4970$  &  $-363.2452$\\ \hline
\multirow{2}*{$D^g$}  &      &      &  $5.872808$  &  $-190.6910$  &    \\ \cline{2-6}  &      &      &  $5.665436$  &  $-179.9663$  &  $1016.432$\\ \hline
\multirow{2}*{$E^g_l$}  &      &  $-7.221632$  &  $30.41847$  &  $-83.58398$  &    \\ \cline{2-6}  &      &  $-7.164407$  &  $30.11041$  &  $-81.90409$  &  $122.4383$\\ \hline
\multirow{2}*{$E^g_h$}  &      &      &  $-7.090949$  &  $55.10048$  &    \\ \cline{2-6}  &      &      &  $-0.2277007$  &  $7.477590$  &  $-38.94969$\\ \hline
\multirow{2}*{$F^g_l$}  &      &  $11.27238$  &  $-51.40444$  &  $145.7744$  &    \\ \cline{2-6}  &      &  $11.11704$  &  $-50.54900$  &  $139.8451$  &  $-187.9518$\\ \hline
\multirow{2}*{$F^g_h$}  &      &      &  $13.78273$  &  $-106.0896$  &    \\ \cline{2-6}  &      &      &      &  $-10.76121$  &  $124.8742$\\ \hline
\multirow{2}*{$G^g_l$}  &      &      &  $13.19603$  &  $-19.72861$  &    \\ \cline{2-6}  &      &      &  $13.04047$  &  $-17.39200$  &  $-19.68276$\\ \hline
\multirow{2}*{$G^g_h$}  &      &      &      &  $18.82390$  &    \\ \cline{2-6}  &      &      &      &      &  $-16.44998$\\ \hline
\multirow{2}*{$H^g_l$}  &      &      &  $1.375549$  &  $-1.554947$  &    \\ \cline{2-6}  &      &      &  $1.364649$  &  $-1.537369$  &  $1.772607$\\ \hline
\multirow{2}*{$H^g_{lh}$}  &      &      &      &  $2.363650$  &    \\ \cline{2-6}  &      &      &      &  $0.07590022$  &  $-1.707018$\\ \hline
\multirow{2}*{$H^g_h$}  &      &      &      &      &    \\ \cline{2-6}  &      &      &      &      &  $0.2450480$\\ \hline
\multirow{2}*{$I^g_l$}  &      &      &  $-2.147121$  &  $2.581333$  &    \\ \cline{2-6}  &      &      &  $-2.117531$  &  $2.532773$  &  $-2.311413$\\ \hline
\multirow{2}*{$I^g_{lh}$}  &      &      &      &  $-4.594245$  &    \\ \cline{2-6}  &      &      &      &      &  $3.483198$\\ \hline
\multirow{2}*{$I^g_h$}  &      &      &      &      &    \\ \cline{2-6}  &      &      &      &      &    \\ \hline
\multirow{2}*{Total}  &  $4.316971$  &  $-34.94825$  &  $146.5483$  &  $-257.9766$  &    \\ \cline{2-6}  &  $4.288068$  &  $-34.65125$  &  $145.6813$  &  $-275.2426$  &  $-381.0666$\\ \hline
Ratio  &  $0.9933$  &  $0.9915$  &  $0.9941$  &  $1.067$  &    \\ \hline

\end{tabular}
\caption{\label{tab:ggnum1} Color-decomposed NNLO squared amplitude in the $gg$ channel at the phase-space point $(\sqrt{s_{12}}, \sqrt{|s_{13}|}, \sqrt{|s_{14}|}, \sqrt{|s_{23}|}, \sqrt{|s_{24}|}) = (2.0000, 1.1041, 1.1373, 1.1648, 1.1402)$~TeV, $m_t=\SI{173}{GeV}$, and $m_H=\SI{125}{GeV}$.}
\end{center}
\end{table}

\begin{table}[t!]
\begin{center}
\begin{tabular}{|l|r|r|r|r|r|}
    \hline
   \text{exact}&\multirow{2}*{$\eps^{-4}$}& \multirow{2}*{$\eps^{-3}$}&\multirow{2}*{$\eps^{-2}$}&\multirow{2}*{$\eps^{-1}$}&\multirow{2}*{$\eps^{0}$}\\ \cline{1-1}
   \text{semi-}&  &  &  &  & \\ \hline

\multirow{2}*{$A^q$}  &  $0.04359554$  &  $-0.4543021$  &  $4.245433$  &  $-28.83733$  &    \\ \cline{2-6}  &  $0.04352110$  &  $-0.4519036$  &  $4.179582$  &  $-28.25268$  &  $129.5765$\\ \hline
\multirow{2}*{$B^q$}  &  $-0.08719108$  &  $0.9328190$  &  $-5.000430$  &  $-4.059032$  &    \\ \cline{2-6}  &  $-0.08704220$  &  $0.9261112$  &  $-4.849566$  &  $10.38085$  &  $41.35127$\\ \hline
\multirow{2}*{$C^q$}  &  $0.04359554$  &  $-0.4785169$  &  $-1.094868$  &  $26.09443$  &    \\ \cline{2-6}  &  $0.04352110$  &  $-0.4742076$  &  $-1.175457$  &  $26.74820$  &  $-129.2190$\\ \hline
\multirow{2}*{$D^q_l$}  &      &  $-0.04359554$  &  $0.01748520$  &  $2.409230$  &    \\ \cline{2-6}  &      &  $-0.04352110$  &  $0.01691567$  &  $2.401585$  &  $-21.19547$\\ \hline
\multirow{2}*{$D^q_h$}  &      &      &  $-0.2852286$  &  $3.073465$  &    \\ \cline{2-6}  &      &      &  $-0.2936883$  &  $3.172621$  &  $-22.32761$\\ \hline
\multirow{2}*{$E^q_l$}  &      &  $0.04359554$  &  $0.05436836$  &  $-0.7929464$  &    \\ \cline{2-6}  &      &  $0.04352110$  &  $0.05543837$  &  $-0.7840151$  &  $-11.00616$\\ \hline
\multirow{2}*{$E^q_h$}  &      &      &  $0.2852286$  &  $-1.225414$  &    \\ \cline{2-6}  &      &      &  $0.2936883$  &  $-1.294804$  &  $-10.45089$\\ \hline
\multirow{2}*{$F^q_l$}  &      &      &      &      &    \\ \cline{2-6}  &      &      &      &      &  $0.6690317$\\ \hline
\multirow{2}*{$F^q_{lh}$}  &      &      &      &      &    \\ \cline{2-6}  &      &      &      &      &  $1.338063$\\ \hline
\multirow{2}*{$F^q_h$}  &      &      &      &      &    \\ \cline{2-6}  &      &      &      &      &  $0.6690317$\\ \hline
\multirow{2}*{Total}  &  $0.1377834$  &  $-1.684597$  &  $14.52404$  &  $-96.47415$  &    \\ \cline{2-6}  &  $0.1375482$  &  $-1.677330$  &  $14.31068$  &  $-87.60797$  &  $360.2332$\\ \hline
Ratio  &  $0.9983$  &  $0.9957$  &  $0.9853$  &  $0.9081$  &    \\ \hline

\end{tabular}
\caption{\label{tab:qqnum2} Color-decomposed NNLO squared amplitude in the $q\bar{q}$ channel at the phase-space point $(\sqrt{s_{12}}, \sqrt{|s_{13}|}, \sqrt{|s_{14}|}, \sqrt{|s_{23}|}, \sqrt{|s_{24}|}) = (5.0000, 1.6558, 3.6246, 3.5411, 2.0659)$~TeV, $m_t=\SI{173}{GeV}$, and $m_H=\SI{125}{GeV}$.}
\end{center}
\end{table}

\begin{table}[t!]
\begin{center}
\begin{tabular}{|l|r|r|r|r|r|}
    \hline
   \text{exact}&\multirow{2}*{$\eps^{-4}$}& \multirow{2}*{$\eps^{-3}$}&\multirow{2}*{$\eps^{-2}$}&\multirow{2}*{$\eps^{-1}$}&\multirow{2}*{$\eps^{0}$}\\ \cline{1-1}
   \text{semi-}&  &  &  &  & \\ \hline

\multirow{2}*{$A^g$}  &  $2.108278$  &  $-22.43687$  &  $154.7362$  &  $-696.5291$  &    \\ \cline{2-6}  &  $2.132335$  &  $-22.66071$  &  $156.2918$  &  $-701.0229$  &  $2001.209$\\ \hline
\multirow{2}*{$B^g$}  &  $-2.321466$  &  $33.50440$  &  $-283.5492$  &  $1587.447$  &    \\ \cline{2-6}  &  $-2.340932$  &  $33.78268$  &  $-286.0874$  &  $1576.900$  &  $-5796.564$\\ \hline
\multirow{2}*{$C^g$}  &      &  $-10.58208$  &  $103.5865$  &  $-310.1786$  &    \\ \cline{2-6}  &      &  $-10.67370$  &  $104.9121$  &  $-522.7946$  &  $1340.553$\\ \hline
\multirow{2}*{$D^g$}  &      &      &  $-6.330904$  &  $8.141297$  &    \\ \cline{2-6}  &      &      &  $-6.390913$  &  $9.683863$  &  $154.2983$\\ \hline
\multirow{2}*{$E^g_l$}  &      &  $-2.459657$  &  $15.63943$  &  $-54.30544$  &    \\ \cline{2-6}  &      &  $-2.487724$  &  $15.81008$  &  $-54.94143$  &  $95.49227$\\ \hline
\multirow{2}*{$E^g_h$}  &      &      &  $-0.4536370$  &  $6.220529$  &    \\ \cline{2-6}  &      &      &  $-0.09288271$  &  $3.065970$  &  $-38.53754$\\ \hline
\multirow{2}*{$F^g_l$}  &      &  $2.708377$  &  $-21.71698$  &  $76.36459$  &    \\ \cline{2-6}  &      &  $2.731087$  &  $-21.90274$  &  $77.13346$  &  $-130.1723$\\ \hline
\multirow{2}*{$F^g_h$}  &      &      &  $0.7115711$  &  $-9.676640$  &    \\ \cline{2-6}  &      &      &      &  $-2.778873$  &  $60.58957$\\ \hline
\multirow{2}*{$G^g_l$}  &      &      &  $5.291039$  &  $-17.18699$  &    \\ \cline{2-6}  &      &      &  $5.336849$  &  $-17.51022$  &  $18.44228$\\ \hline
\multirow{2}*{$G^g_h$}  &      &      &      &  $1.621799$  &    \\ \cline{2-6}  &      &      &      &      &  $-12.54769$\\ \hline
\multirow{2}*{$H^g_l$}  &      &      &  $0.4685062$  &  $-0.7416234$  &    \\ \cline{2-6}  &      &      &  $0.4738522$  &  $-0.7528461$  &  $0.7176704$\\ \hline
\multirow{2}*{$H^g_{lh}$}  &      &      &      &  $0.1512123$  &    \\ \cline{2-6}  &      &      &      &  $0.03096090$  &  $-0.5283240$\\ \hline
\multirow{2}*{$H^g_h$}  &      &      &      &      &    \\ \cline{2-6}  &      &      &      &      &  $0.1612425$\\ \hline
\multirow{2}*{$I^g_l$}  &      &      &  $-0.5158813$  &  $0.8488621$  &    \\ \cline{2-6}  &      &      &  $-0.5202071$  &  $0.8576413$  &  $-0.6890958$\\ \hline
\multirow{2}*{$I^g_{lh}$}  &      &      &      &  $-0.2371904$  &    \\ \cline{2-6}  &      &      &      &      &  $0.8557063$\\ \hline
\multirow{2}*{$I^g_h$}  &      &      &      &      &    \\ \cline{2-6}  &      &      &      &      &    \\ \hline
\multirow{2}*{Total}  &  $1.561222$  &  $-18.93601$  &  $124.5975$  &  $-506.8341$  &    \\ \cline{2-6}  &  $1.579695$  &  $-19.13566$  &  $125.9882$  &  $-515.3526$  &  $1265.827$\\ \hline
Ratio  &  $1.012$  &  $1.011$  &  $1.011$  &  $1.017$  &    \\ \hline

\end{tabular}
\caption{\label{tab:ggnum2} Color-decomposed NNLO squared amplitude in the $gg$ channel at the phase-space point $(\sqrt{s_{12}}, \sqrt{|s_{13}|}, \sqrt{|s_{14}|}, \sqrt{|s_{23}|}, \sqrt{|s_{24}|}) = (5.0000, 1.6558, 3.6246, 3.5411, 2.0659)$~TeV, $m_t=\SI{173}{GeV}$, and $m_H=\SI{125}{GeV}$.}
\end{center}
\end{table}

\begin{table}[t!]
\begin{center}
\begin{tabular}{|l|r|r|r|r|r|}
    \hline
   \text{exact}&\multirow{2}*{$\eps^{-4}$}& \multirow{2}*{$\eps^{-3}$}&\multirow{2}*{$\eps^{-2}$}&\multirow{2}*{$\eps^{-1}$}&\multirow{2}*{$\eps^{0}$}\\ \cline{1-1}
   \text{semi-}&  &  &  &  & \\ \hline
   
\multirow{2}*{$A^q$}  &  $0.01036634$  &  $-0.1657161$  &  $1.867325$  &  $-14.63674$  &    \\ \cline{2-6}  &  $0.01036192$  &  $-0.1655484$  &  $1.862399$  &  $-14.58049$  &  $77.26753$\\ \hline
\multirow{2}*{$B^q$}  &  $-0.02073268$  &  $0.3368775$  &  $-2.685072$  &  $6.796855$  &    \\ \cline{2-6}  &  $-0.02072384$  &  $0.3364287$  &  $-2.674131$  &  $11.54460$  &  $-14.12207$\\ \hline
\multirow{2}*{$C^q$}  &  $0.01036634$  &  $-0.1711614$  &  $0.3780909$  &  $5.791447$  &    \\ \cline{2-6}  &  $0.01036192$  &  $-0.1708803$  &  $0.3723369$  &  $5.854854$  &  $-49.48764$\\ \hline
\multirow{2}*{$D^q_l$}  &      &  $-0.01036634$  &  $0.004553374$  &  $1.032195$  &    \\ \cline{2-6}  &      &  $-0.01036192$  &  $0.004519188$  &  $1.031455$  &  $-10.54324$\\ \hline
\multirow{2}*{$D^q_h$}  &      &      &  $-0.08854632$  &  $1.279040$  &    \\ \cline{2-6}  &      &      &  $-0.08907700$  &  $1.287382$  &  $-10.97078$\\ \hline
\multirow{2}*{$E^q_l$}  &      &  $0.01036634$  &  $0.01263650$  &  $-0.6003433$  &    \\ \cline{2-6}  &      &  $0.01036192$  &  $0.01270037$  &  $-0.5994460$  &  $-0.6874765$\\ \hline
\multirow{2}*{$E^q_h$}  &      &      &  $0.08854632$  &  $-0.7862173$  &    \\ \cline{2-6}  &      &      &  $0.08907700$  &  $-0.7926235$  &  $-0.3985877$\\ \hline
\multirow{2}*{$F^q_l$}  &      &      &      &      &    \\ \cline{2-6}  &      &      &      &      &  $0.2919736$\\ \hline
\multirow{2}*{$F^q_{lh}$}  &      &      &      &      &    \\ \cline{2-6}  &      &      &      &      &  $0.5839472$\\ \hline
\multirow{2}*{$F^q_h$}  &      &      &      &      &    \\ \cline{2-6}  &      &      &      &      &  $0.2919736$\\ \hline
\multirow{2}*{Total}  &  $0.03276276$  &  $-0.5830237$  &  $6.229379$  &  $-47.21463$  &    \\ \cline{2-6}  &  $0.03274878$  &  $-0.5825123$  &  $6.213445$  &  $-44.87048$  &  $219.5371$\\ \hline
Ratio  &  $0.9996$  &  $0.9991$  &  $0.9974$  &  $0.9504$  &    \\ \hline

\end{tabular}
\caption{Color-decomposed NNLO squared amplitude in the $q\bar{q}$ channel at the phase-space point $(\sqrt{s_{12}}, \sqrt{|s_{13}|}, \sqrt{|s_{14}|}, \sqrt{|s_{23}|}, \sqrt{|s_{24}|}) = (10.000, 3.3073, 7.2580, 7.0877, 4.1365)$~TeV, $m_t=\SI{173}{GeV}$, and $m_H=\SI{125}{GeV}$.}
\label{tab:qqnum3}
\end{center}
\end{table}

\begin{table}[t!]
\begin{center}
\begin{tabular}{|l|r|r|r|r|r|}
    \hline
   \text{exact}&\multirow{2}*{$\eps^{-4}$}& \multirow{2}*{$\eps^{-3}$}&\multirow{2}*{$\eps^{-2}$}&\multirow{2}*{$\eps^{-1}$}&\multirow{2}*{$\eps^{0}$}\\ \cline{1-1}
   \text{semi-}&  &  &  &  & \\ \hline
   
\multirow{2}*{$A^g$}  &  $0.5428451$  &  $-8.014744$  &  $70.73525$  &  $-409.1486$  &    \\ \cline{2-6}  &  $0.5444174$  &  $-8.035900$  &  $70.93871$  &  $-410.1906$  &  $1595.269$\\ \hline
\multirow{2}*{$B^g$}  &  $-0.5947800$  &  $11.80785$  &  $-122.2304$  &  $816.5623$  &    \\ \cline{2-6}  &  $-0.5960768$  &  $11.83389$  &  $-122.5573$  &  $812.4284$  &  $-3618.243$\\ \hline
\multirow{2}*{$C^g$}  &      &  $-3.535467$  &  $46.32135$  &  $-235.2084$  &    \\ \cline{2-6}  &      &  $-3.543373$  &  $46.48641$  &  $-306.0970$  &  $1153.332$\\ \hline
\multirow{2}*{$D^g$}  &      &      &  $-3.786279$  &  $20.42499$  &    \\ \cline{2-6}  &      &      &  $-3.795128$  &  $20.68441$  &  $-16.55695$\\ \hline
\multirow{2}*{$E^g_l$}  &      &  $-0.6333192$  &  $5.140016$  &  $-20.91982$  &    \\ \cline{2-6}  &      &  $-0.6351536$  &  $5.154407$  &  $-20.98816$  &  $40.29084$\\ \hline
\multirow{2}*{$E^g_h$}  &      &      &  $-0.05982249$  &  $1.273500$  &    \\ \cline{2-6}  &      &      &  $-0.02446752$  &  $0.8755931$  &  $-18.82327$\\ \hline
\multirow{2}*{$F^g_l$}  &      &  $0.6939100$  &  $-7.171378$  &  $29.55495$  &    \\ \cline{2-6}  &      &  $0.6954229$  &  $-7.187486$  &  $29.64513$  &  $-56.16316$\\ \hline
\multirow{2}*{$F^g_h$}  &      &      &  $0.07058572$  &  $-1.610635$  &    \\ \cline{2-6}  &      &      &      &  $-0.7263948$  &  $27.30988$\\ \hline
\multirow{2}*{$G^g_l$}  &      &      &  $1.767734$  &  $-7.027765$  &    \\ \cline{2-6}  &      &      &  $1.771687$  &  $-7.066248$  &  $10.48010$\\ \hline
\multirow{2}*{$G^g_h$}  &      &      &      &  $0.2097864$  &    \\ \cline{2-6}  &      &      &      &      &  $-5.628954$\\ \hline
\multirow{2}*{$H^g_l$}  &      &      &  $0.1206322$  &  $-0.1890601$  &    \\ \cline{2-6}  &      &      &  $0.1209816$  &  $-0.1897854$  &  $0.1810913$\\ \hline
\multirow{2}*{$H^g_{lh}$}  &      &      &      &  $0.01994083$  &    \\ \cline{2-6}  &      &      &      &  $0.008155841$  &  $-0.1244663$\\ \hline
\multirow{2}*{$H^g_h$}  &      &      &      &      &    \\ \cline{2-6}  &      &      &      &      &  $0.05258422$\\ \hline
\multirow{2}*{$I^g_l$}  &      &      &  $-0.1321733$  &  $0.2159630$  &    \\ \cline{2-6}  &      &      &  $-0.1324615$  &  $0.2165426$  &  $-0.1742280$\\ \hline
\multirow{2}*{$I^g_{lh}$}  &      &      &      &  $-0.02352857$  &    \\ \cline{2-6}  &      &      &      &      &  $0.2178904$\\ \hline
\multirow{2}*{$I^g_h$}  &      &      &      &      &    \\ \cline{2-6}  &      &      &      &      &    \\ \hline
\multirow{2}*{Total}  &  $0.4022649$  &  $-6.474464$  &  $55.07390$  &  $-296.0855$  &    \\ \cline{2-6}  &  $0.4034700$  &  $-6.492298$  &  $55.24290$  &  $-298.2640$  &  $1062.735$\\ \hline
Ratio  &  $1.003$  &  $1.003$  &  $1.003$  &  $1.007$  &    \\ \hline

\end{tabular}
\caption{Color-decomposed NNLO squared amplitude in the $gg$ channel at the phase-space point $(\sqrt{s_{12}}, \sqrt{|s_{13}|}, \sqrt{|s_{14}|}, \sqrt{|s_{23}|}, \sqrt{|s_{24}|}) = (10.000, 3.3073, 7.2580, 7.0877, 4.1365)$~TeV, $m_t=\SI{173}{GeV}$, and $m_H=\SI{125}{GeV}$.}
\label{tab:ggnum3}
\end{center}
\end{table}

\clearpage


\appendix

\section{Spin (Lorentz) structures}\label{sec:spinstructures}

\subsection{Massive spin (Lorentz) structures}\label{sec:massivebases}

For the massive $q\bar{q}$ channel, we define $\ket{d_i^q}$ in the following form:
\begin{align}
\ket{d_i^q} = \bar{v}(p_2) \, \Gamma_i \, u(p_1) \, v(p_3) \, \Gamma'_i \, \bar{u}(p_4) \quad \Rightarrow \quad \mC{S}_i^q = \Gamma_i \otimes \Gamma'_i \,, 
\end{align}
where $\Gamma_i$ denotes a string of $\gamma$ matrices concerning the initial state fermion line, while $\Gamma'_i$ concerns the final state fermion line. $\mC{S}_i^q$ are given by
\begin{align}
 \mC{S}_{1}^q &= \gamma^{\mu} \otimes \gamma^{\mu}\,, &
 \mC{S}_{2}^q &= \gamma^{\mu} \otimes \gamma^{\mu}  \slashed{p}_1\,, \nn \\ 
 \mC{S}_{3}^q &= \gamma^{\mu} \otimes \gamma^{\mu}  \slashed{p}_2\,, &
 \mC{S}_{4}^q &= \gamma^{\mu} \otimes \gamma^{\mu}  \slashed{p}_1  \slashed{p}_2\,, \nn \\ 
 \mC{S}_{5}^q &= \gamma^{\mu}  \slashed{p}_3  \slashed{p}_4 \otimes \gamma^{\mu}\,, &
 \mC{S}_{6}^q &= \gamma^{\mu}  \slashed{p}_3  \slashed{p}_4 \otimes \gamma^{\mu}  \slashed{p}_1\,, \nn \\ 
 \mC{S}_{7}^q &= \gamma^{\mu}  \slashed{p}_3  \slashed{p}_4 \otimes \gamma^{\mu}  \slashed{p}_2\,, &
 \mC{S}_{8}^q &= \gamma^{\mu}  \slashed{p}_3  \slashed{p}_4 \otimes \gamma^{\mu}  \slashed{p}_1  \slashed{p}_2\,, \nn \\ 
 \mC{S}_{9}^q &= \slashed{p}_3  \otimes \mathbf{1}\,, &
 \mC{S}_{10}^q &= \slashed{p}_3 \otimes \slashed{p}_1\,, \nn \\ 
 \mC{S}_{11}^q &= \slashed{p}_3 \otimes \slashed{p}_2\,, &
 \mC{S}_{12}^q &= \slashed{p}_3 \otimes \slashed{p}_1  \slashed{p}_2\,, \nn \\ 
 \mC{S}_{13}^q &= \slashed{p}_4  \otimes \mathbf{1}\,, &
 \mC{S}_{14}^q &= \slashed{p}_4 \otimes \slashed{p}_1\,, \nn \\ 
 \mC{S}_{15}^q &= \slashed{p}_4 \otimes \slashed{p}_2\,, &
 \mC{S}_{16}^q &= \slashed{p}_4 \otimes \slashed{p}_1  \slashed{p}_2\,, \nn \\ 
 \mC{S}_{17}^q &= \gamma^{\mu}  \gamma^{\nu}  \slashed{p}_3 \otimes \gamma^{\mu}  \gamma^{\nu}\,, &
 \mC{S}_{18}^q &= \gamma^{\mu}  \gamma^{\nu}  \slashed{p}_3 \otimes \gamma^{\mu}  \gamma^{\nu}  \slashed{p}_1\,, \nn \\ 
 \mC{S}_{19}^q &= \gamma^{\mu}  \gamma^{\nu}  \slashed{p}_3 \otimes \gamma^{\mu}  \gamma^{\nu}  \slashed{p}_2\,, &
 \mC{S}_{20}^q &= \gamma^{\mu}  \gamma^{\nu}  \slashed{p}_3 \otimes \gamma^{\mu}  \gamma^{\nu}  \slashed{p}_1  \slashed{p}_2\,, \nn \\ 
 \mC{S}_{21}^q &= \gamma^{\mu}  \gamma^{\nu}  \slashed{p}_4 \otimes \gamma^{\mu}  \gamma^{\nu}\,, &
 \mC{S}_{22}^q &= \gamma^{\mu}  \gamma^{\nu}  \slashed{p}_4 \otimes \gamma^{\mu}  \gamma^{\nu}  \slashed{p}_1\,, \nn \\ 
 \mC{S}_{23}^q &= \gamma^{\mu}  \gamma^{\nu}  \slashed{p}_4 \otimes \gamma^{\mu}  \gamma^{\nu}  \slashed{p}_2\,, &
 \mC{S}_{24}^q &= \gamma^{\mu}  \gamma^{\nu}  \slashed{p}_4 \otimes \gamma^{\mu}  \gamma^{\nu}  \slashed{p}_1  \slashed{p}_2\,, \nn \\ 
 \mC{S}_{25}^q &= \gamma^{\mu}  \gamma^{\nu}  \gamma^{\rho} \otimes \gamma^{\mu}  \gamma^{\nu}  \gamma^{\rho}\,, &
 \mC{S}_{26}^q &= \gamma^{\mu}  \gamma^{\nu}  \gamma^{\rho} \otimes \gamma^{\mu}  \gamma^{\nu}  \gamma^{\rho}  \slashed{p}_1\,, \nn \\ 
 \mC{S}_{27}^q &= \gamma^{\mu}  \gamma^{\nu}  \gamma^{\rho} \otimes \gamma^{\mu}  \gamma^{\nu}  \gamma^{\rho}  \slashed{p}_2\,, &
 \mC{S}_{28}^q &= \gamma^{\mu}  \gamma^{\nu}  \gamma^{\rho} \otimes \gamma^{\mu}  \gamma^{\nu}  \gamma^{\rho}  \slashed{p}_1  \slashed{p}_2 \,.
\end{align}

For the massive $gg$ channel, we define $\ket{d_i^g}$ in the following form:
\begin{align}
\ket{d_i^g} = \varepsilon_{\mu}\lt(p_1,s\rt)\varepsilon_{\nu}\lt(p_2,s\rt) \, v(p_3) \, \Gamma^{\mu\nu}_i \, \bar{u}(p_4) \,, 
\end{align}
where $\Gamma^{\mu\nu}_i$ are given by
\begin{align}
 \Gamma^{\mu\nu}_{1} &= \gamma^{\mu}  \gamma^{\nu}  \slashed{p}_1  \slashed{p}_2\,, &
 \Gamma^{\mu\nu}_{2} &= \mathbf{1} g^{\mu\nu}\,, &
 \Gamma^{\mu\nu}_{3} &= \mathbf{1} p_3^{\mu} p_3^{\nu}\,, &
 \Gamma^{\mu\nu}_{4} &= \mathbf{1} p_4^{\mu} p_4^{\nu}\,, \nn \\ 
 \Gamma^{\mu\nu}_{5} &= \gamma^{\mu}  \gamma^{\nu}\,, &
 \Gamma^{\mu\nu}_{6} &= \slashed{p}_1  \slashed{p}_2 g^{\mu\nu}\,, &
 \Gamma^{\mu\nu}_{7} &= \slashed{p}_1  \slashed{p}_2 p_3^{\mu} p_3^{\nu}\,, &
 \Gamma^{\mu\nu}_{8} &= \slashed{p}_1  \slashed{p}_2 p_4^{\mu} p_4^{\nu}\,, \nn \\ 
 \Gamma^{\mu\nu}_{9} &= \gamma^{\mu}  \gamma^{\nu}  \slashed{p}_1\,, &
 \Gamma^{\mu\nu}_{10} &= \gamma^{\mu}  \slashed{p}_1  \slashed{p}_2 p_3^{\nu}\,, &
 \Gamma^{\mu\nu}_{11} &= \gamma^{\mu}  \slashed{p}_1  \slashed{p}_2 p_4^{\nu}\,, &
 \Gamma^{\mu\nu}_{12} &= \gamma^{\mu}  \slashed{p}_1 p_3^{\nu}\,, \nn \\ 
 \Gamma^{\mu\nu}_{13} &= \gamma^{\mu}  \slashed{p}_1 p_4^{\nu}\,, &
 \Gamma^{\mu\nu}_{14} &= \gamma^{\mu}  \slashed{p}_2 p_3^{\nu}\,, &
 \Gamma^{\mu\nu}_{15} &= \gamma^{\mu}  \slashed{p}_2 p_4^{\nu}\,, &
 \Gamma^{\mu\nu}_{16} &= \slashed{p}_1  \slashed{p}_2 p_3^{\mu} p_4^{\nu}\,, \nn \\ 
 \Gamma^{\mu\nu}_{17} &= \gamma^{\mu} p_3^{\nu}\,, &
 \Gamma^{\mu\nu}_{18} &= \gamma^{\mu} p_4^{\nu}\,, &
 \Gamma^{\mu\nu}_{19} &= \slashed{p}_1 g^{\mu\nu}\,, &
 \Gamma^{\mu\nu}_{20} &= \slashed{p}_1 p_3^{\mu} p_3^{\nu}\,, \nn \\ 
 \Gamma^{\mu\nu}_{21} &= \slashed{p}_1 p_3^{\mu} p_4^{\nu}\,, &
 \Gamma^{\mu\nu}_{22} &= \slashed{p}_1 p_3^{\nu} p_4^{\mu}\,, &
 \Gamma^{\mu\nu}_{23} &= \slashed{p}_1 p_4^{\mu} p_4^{\nu}\,, &
 \Gamma^{\mu\nu}_{24} &= \mathbf{1} p_3^{\mu} p_4^{\nu}\,, \nn \\ 
 \Gamma^{\mu\nu}_{25} &= \gamma^{\mu}  \gamma^{\nu}  \slashed{p}_2\,, &
 \Gamma^{\mu\nu}_{26} &= \gamma^{\nu}  \slashed{p}_1  \slashed{p}_2 p_3^{\mu}\,, &
 \Gamma^{\mu\nu}_{27} &= \gamma^{\nu}  \slashed{p}_1  \slashed{p}_2 p_4^{\mu}\,, &
 \Gamma^{\mu\nu}_{28} &= \gamma^{\nu}  \slashed{p}_2 p_3^{\mu}\,, \nn \\ 
 \Gamma^{\mu\nu}_{29} &= \gamma^{\nu}  \slashed{p}_2 p_4^{\mu}\,, &
 \Gamma^{\mu\nu}_{30} &= \gamma^{\nu}  \slashed{p}_1 p_3^{\mu}\,, &
 \Gamma^{\mu\nu}_{31} &= \gamma^{\nu}  \slashed{p}_1 p_4^{\mu}\,, &
 \Gamma^{\mu\nu}_{32} &= \slashed{p}_1  \slashed{p}_2 p_3^{\nu} p_4^{\mu}\,, \nn \\ 
 \Gamma^{\mu\nu}_{33} &= \gamma^{\nu} p_3^{\mu}\,, &
 \Gamma^{\mu\nu}_{34} &= \gamma^{\nu} p_4^{\mu}\,, &
 \Gamma^{\mu\nu}_{35} &= \slashed{p}_2 g^{\mu\nu}\,, &
 \Gamma^{\mu\nu}_{36} &= \slashed{p}_2 p_3^{\mu} p_3^{\nu}\,, \nn \\ 
 \Gamma^{\mu\nu}_{37} &= \slashed{p}_2 p_3^{\nu} p_4^{\mu}\,, &
 \Gamma^{\mu\nu}_{38} &= \slashed{p}_2 p_3^{\mu} p_4^{\nu}\,, &
 \Gamma^{\mu\nu}_{39} &= \slashed{p}_2 p_4^{\mu} p_4^{\nu}\,, &
 \Gamma^{\mu\nu}_{40} &= \mathbf{1} p_3^{\nu} p_4^{\mu} \,.
\end{align}

\subsection{Massless spin (Lorentz) structures}\label{sec:masslessbases}
For the massless $q\bar{q}$ channel, we define $\ket{\td{d}_i^q}$ in the following form:
\begin{align}
\ket{\td{d}_i^q} = \bar{v}(\td{p}_2) \, \td{\Gamma}_i \, u(\td{p}_1) \, v(\td{p}_3) \, \td{\Gamma}'_i \, \bar{u}(\td{p}_4) \quad \Rightarrow \quad \td{\mC{S}}_i^q = \td{\Gamma}_i \otimes \td{\Gamma}'_i \,,
\end{align}
where $\td{\Gamma}_i$ denotes a string of $\gamma$ matrices concerning the initial state fermion line, while $\td{\Gamma}'_i$ concerns the final state fermion line. $\td{\mC{S}}_i^q$ are given by
\begin{align}
 \td{\mC{S}}_{1}^q &= \td{\slashed{p}}_3  \otimes \mathbf{1}\,, &
 \td{\mC{S}}_{2}^q &= \td{\slashed{p}}_4  \otimes \mathbf{1}\,, \nn \\ 
 \td{\mC{S}}_{3}^q &= \gamma^{\mu} \otimes \gamma^{\mu}  \td{\slashed{p}}_1\,, &
 \td{\mC{S}}_{4}^q &= \gamma^{\mu} \otimes \gamma^{\mu}  \td{\slashed{p}}_2\,, \nn \\ 
 \td{\mC{S}}_{5}^q &= \td{\slashed{p}}_3 \otimes \td{\slashed{p}}_1  \td{\slashed{p}}_2\,, &
 \td{\mC{S}}_{6}^q &= \td{\slashed{p}}_4 \otimes \td{\slashed{p}}_1  \td{\slashed{p}}_2\,, \nn \\ 
 \td{\mC{S}}_{7}^q &= \gamma^{\mu}  \gamma^{\nu}  \td{\slashed{p}}_3 \otimes \gamma^{\mu}  \gamma^{\nu}\,, &
 \td{\mC{S}}_{8}^q &= \gamma^{\mu}  \gamma^{\nu}  \td{\slashed{p}}_4 \otimes \gamma^{\mu}  \gamma^{\nu}\,, \nn \\ 
 \td{\mC{S}}_{9}^q &= \gamma^{\mu}  \td{\slashed{p}}_3  \td{\slashed{p}}_4 \otimes \gamma^{\mu}  \td{\slashed{p}}_1\,, &
 \td{\mC{S}}_{10}^q &= \gamma^{\mu}  \td{\slashed{p}}_3  \td{\slashed{p}}_4 \otimes \gamma^{\mu}  \td{\slashed{p}}_2\,, \nn \\ 
 \td{\mC{S}}_{11}^q &= \gamma^{\mu}  \gamma^{\nu}  \gamma^{\rho} \otimes \gamma^{\mu}  \gamma^{\nu}  \gamma^{\rho}  \td{\slashed{p}}_1\,, &
 \td{\mC{S}}_{12}^q &= \gamma^{\mu}  \gamma^{\nu}  \gamma^{\rho} \otimes \gamma^{\mu}  \gamma^{\nu}  \gamma^{\rho}  \td{\slashed{p}}_2\,, \nn \\ 
 \td{\mC{S}}_{13}^q &= \gamma^{\mu}  \gamma^{\nu}  \td{\slashed{p}}_3 \otimes \gamma^{\mu}  \gamma^{\nu}  \td{\slashed{p}}_1  \td{\slashed{p}}_2\,, &
 \td{\mC{S}}_{14}^q &= \gamma^{\mu}  \gamma^{\nu}  \td{\slashed{p}}_4 \otimes \gamma^{\mu}  \gamma^{\nu}  \td{\slashed{p}}_1  \td{\slashed{p}}_2\,.
\end{align}

For the massless $gg$ channel, we define $\ket{\td{d}_i^g}$ in the following form:
\begin{align}
\ket{\td{d}_i^g} = \varepsilon_{\mu}\lt(\td{p}_1,s\rt)\varepsilon_{\nu}\lt(\td{p}_2,s\rt) \, v(\td{p}_3) \, \td{\Gamma}^{\mu\nu}_i \, \bar{u}(\td{p}_4) \,, 
\end{align}
where $\td{\Gamma}^{\mu\nu}_i$ are given by
\begin{align}
 \td{\Gamma}^{\mu\nu}_{1} &= \gamma^{\mu}  \gamma^{\nu}\,, &
 \td{\Gamma}^{\mu\nu}_{2} &= \gamma^{\mu}  \gamma^{\nu}  \td{\slashed{p}}_1  \td{\slashed{p}}_2\,, &
 \td{\Gamma}^{\mu\nu}_{3} &= \mathbf{1} g^{\mu\nu}\,, &
 \td{\Gamma}^{\mu\nu}_{4} &= \td{\slashed{p}}_1  \td{\slashed{p}}_2 g^{\mu\nu}\,, \nn \\ 
 \td{\Gamma}^{\mu\nu}_{5} &= \gamma^{\nu}  \td{\slashed{p}}_1 \td{p}_3^{\mu}\,, &
 \td{\Gamma}^{\mu\nu}_{6} &= \gamma^{\nu}  \td{\slashed{p}}_2 \td{p}_3^{\mu}\,, &
 \td{\Gamma}^{\mu\nu}_{7} &= \gamma^{\nu}  \td{\slashed{p}}_1 \td{p}_4^{\mu}\,, &
 \td{\Gamma}^{\mu\nu}_{8} &= \gamma^{\nu}  \td{\slashed{p}}_2 \td{p}_4^{\mu}\,, \nn \\ 
 \td{\Gamma}^{\mu\nu}_{9} &= \gamma^{\mu}  \td{\slashed{p}}_1 \td{p}_3^{\nu}\,, &
 \td{\Gamma}^{\mu\nu}_{10} &= \gamma^{\mu}  \td{\slashed{p}}_2 \td{p}_3^{\nu}\,, &
 \td{\Gamma}^{\mu\nu}_{11} &= \mathbf{1} \td{p}_3^{\mu} \td{p}_3^{\nu}\,, &
 \td{\Gamma}^{\mu\nu}_{12} &= \td{\slashed{p}}_1  \td{\slashed{p}}_2 \td{p}_3^{\mu} \td{p}_3^{\nu}\,, \nn \\ 
 \td{\Gamma}^{\mu\nu}_{13} &= \mathbf{1} \td{p}_3^{\nu} \td{p}_4^{\mu}\,, &
 \td{\Gamma}^{\mu\nu}_{14} &= \td{\slashed{p}}_1  \td{\slashed{p}}_2 \td{p}_3^{\nu} \td{p}_4^{\mu}\,, &
 \td{\Gamma}^{\mu\nu}_{15} &= \gamma^{\mu}  \td{\slashed{p}}_1 \td{p}_4^{\nu}\,, &
 \td{\Gamma}^{\mu\nu}_{16} &= \gamma^{\mu}  \td{\slashed{p}}_2 \td{p}_4^{\nu}\,, \nn \\ 
 \td{\Gamma}^{\mu\nu}_{17} &= \mathbf{1} \td{p}_3^{\mu} \td{p}_4^{\nu}\,, &
 \td{\Gamma}^{\mu\nu}_{18} &= \td{\slashed{p}}_1  \td{\slashed{p}}_2 \td{p}_3^{\mu} \td{p}_4^{\nu}\,, &
 \td{\Gamma}^{\mu\nu}_{19} &= \mathbf{1} \td{p}_4^{\mu} \td{p}_4^{\nu}\,, &
 \td{\Gamma}^{\mu\nu}_{20} &= \td{\slashed{p}}_1  \td{\slashed{p}}_2 \td{p}_4^{\mu} \td{p}_4^{\nu}\,.
\end{align}

\section{Renormalization constant}\label{sec:renormalconstant}
In this section, we present the various renormalization constants needed in our work. Up to NNLO, the renormalization constant of QCD coupling $\alpha_s$ is given by 
\begin{equation}
Z_{\alpha_s}=1-\lt(\fc{\alpha_s}{4\pi}\rt)\fc{\beta_0}{\eps}+\lt(\fc{\alpha_s}{4\pi}\rt)^2\lt(\fc{\beta_0^2}{\eps^2}-\fc{\beta_1}{2\eps}\rt) \,,
\end{equation}
where
\begin{equation}
\beta_0=\fc{11}{3}C_A-\fc{4}{3}T_Fn_f,\quad \beta_1=\fc{34}{3}C_A^2-\fc{20}{3}C_AT_Fn_f-4C_FT_Fn_f.
\end{equation}
Up to NNLO, the top-quark mass renormalization constant in the on-shell scheme is given by: 
\begin{align}
Z_m &= 1 + \frac{\alpha_s}{4\pi} C_F \left[ -\frac{3}{\epsilon} - \left(4+3L_\mu\right) - \epsilon\left(8+4L_\mu+\frac{\pi^2}{4}+\frac{3}{2}L_\mu^2\right) \right. \\ & \left.+\epsilon^2\left(-L_{\mu}(8+\frac{\pi^2}{4}) -2L_{\mu}^2-\frac{L_{\mu}^3}{2} -48-\frac{\pi^2}{3}+\zeta_3 \right)\right]  \nonumber \\
&+ \left(\frac{\alpha_s}{4\pi}\right)^2 \bigg\{C_A \left[\frac{11}{2 \epsilon ^2}-\frac{97}{12 \epsilon
   }+\frac{4 \pi ^2}{3}-\frac{1111}{24}+6 \zeta_3-\frac{11 L_{\mu }^2}{2} -4 \pi ^2 L_{\mu }-\frac{185 L_{\mu }}{6}\right]\nonumber \\
&+C_F \left[\frac{9}{2 \epsilon ^2}+\frac{36 L_{\mu
   }+45}{4\epsilon } +\frac{1}{8} \left(-96 \zeta_3+199-34 \pi ^2\right) + 9 L_{\mu }^2+8 \pi ^2 L_{\mu } +\frac{45 L_{\mu }}{2}\right]\nonumber \\
& + T_Fn_h
   \bigg(-\frac{2}{\epsilon ^2}+\frac{5}{3 \epsilon } +2 L_{\mu }^2+\frac{26 L_{\mu }}{3}-\frac{8 \pi ^2}{3}+\frac{143}{6}\bigg)\nonumber \\
&+T_Fn_l
   \bigg[-\frac{2}{\epsilon ^2}+\frac{5}{3 \epsilon } +\frac{1}{6} \left(71+8 \pi ^2\right)+2 L_{\mu }^2+\frac{26 L_{\mu }}{3}\bigg]\bigg\}+ \mathcal{O}(\alpha_s^3) \, ,
\end{align}
which is also used to renormalize the top-Higgs Yuakawa coupling according to Eq.~\eqref{eq:renorcoupling}. The on-shell wave-function renormalization constants are
\begin{align}
Z_{q} &= 1 + \mathcal{O}(\alpha_s^2) \, , \nonumber
\\
Z_{g} &= 1 + \frac{\alpha_s}{4\pi} T_Fn_h \left[ -\frac{4}{3\epsilon} - \frac{4}{3}L_{\mu} - \epsilon\left(\frac{\pi^2}{9} + \frac{2}{3}L^2_{\mu} \right)+\epsilon^2\left( -\frac{\pi^2}{9} L_{\mu}-\frac{2}{9} L_{\mu}^3 +\frac{4}{9}\zeta_3\right)\right]\\ & 
 \qquad+ \mathcal{O}(\alpha_s^2) \, , \nonumber
\\
Z_{Q} &= 1 + \frac{\alpha_s}{4\pi} C_F \left[ -\frac{3}{\epsilon} - \left(4+3L_\mu\right) - \epsilon\left(8+4L_\mu+\frac{\pi^2}{4}+\frac{3}{2}L_\mu^2\right) \right. \\ & 
\qquad\left.+\epsilon^2\left(-L_{\mu}(8+\frac{\pi^2}{4}) -2L_{\mu}^2-\frac{L_{\mu}^3}{2} -48-\frac{\pi^2}{3}+\zeta_3 \right) \right] + \mathcal{O}(\alpha_s^2) \, .
\end{align}

\section{Matching coefficient in mass-factorization formula}\label{sec:mczfac}
In this section, we present the soft function $\bm{\scal}$ and the $\mathcal{Z}$-factors $\zcal_{[i]}^{(m|0)}$ for gluons and quarks up to NNLO , which can be expanded in strong coupling constant $\alpha_s$ as
\begin{equation}
\begin{aligned}
\bm{\scal}(\{p\},\{m\},\epsilon)&=1+ \sum_{n=1}^\infty \left( \frac{\alpha_s}{4\pi} \right)^n \, \bm{\scal}^{(n)} \,,\\
\zcal_{[j]}^{(m|0)} (\{m\},\epsilon) &= 1 + \sum_{n=1}^\infty \left( \frac{\alpha_s}{4\pi} \right)^n \, \zcal_{[j]}^{(n)} \,.
\end{aligned}
\end{equation}
General results of soft function and $\zcal$-factors can be found in \cite{Mitov:2006xs, Czakon:2007ej, Czakon:2007wk,Becher:2007cu, Engel:2018fsb, Wang:2023qbf}, here we only present the results satisfied for the $t\bar t H$ production process.
Up to the second order, the soft function is given by
\begin{equation}
\bm{\scal}(\{\tilde{p}\},m_t) = 1 + \left( \frac{\alpha_s}{4\pi} \right)^2 \sum_{\substack{i,j\\i \neq j}} \left(-\bm{T}_i \cdot \bm{T}_j\right)  \mathcal{S}^{(2)}(\tilde{s}_{ij},m_t^2) + \mathcal{O}(\alpha_s^3) \,,
\end{equation}
where $i$ and $j$ run over all colored external legs, and
\begin{align}\label{eq:softfunc}
\scal^{(2)}(\tilde{s}_{ij},m_t^2) &= T_F \left( \frac{\mu^2}{m_t^2} \right)^{2\epsilon} \left( -\frac{4}{3\epsilon^2} + \frac{20}{9\epsilon} - \frac{112}{27} - \frac{4\zeta_2}{3} \right) \ln\frac{-\tilde{s}_{ij}}{m^2_{t}} \,.
\end{align}
The boldface $\bm{T}_i$ is the color generator for the external parton $i$ which is an operator in the color space \cite{Catani:1996jh, Catani:1996vz}. For a final-state quark or an initial-state anti-quark, $(\bm{T}^a_i)_{\alpha\beta} = t^a_{\alpha\beta}$; for a final-state anti-quark or an initial-state quark, $(\bm{T}^a_i)_{\alpha\beta} = -t^a_{\beta\alpha}$; and for a gluon, $(\bm{T}^a_i)_{bc} = -i f^{abc}$. The dot product is $\bm{T}_i \cdot \bm{T}_j \equiv \bm{T}_i^a \bm{T}_j^a$ with repeated indices summed over.

The $\zcal$-factor for massless quarks starts at $\alpha_s^2$, and is given by
\begin{equation}
\zcal_{[q]}^{(2)} =  C_F T_F \left( \frac{\mu^2}{m_t^2} \right)^{2\epsilon} \left[ \frac{2}{\epsilon^3} + \frac{8}{9\epsilon^2} - \frac{1}{\epsilon} \left( \frac{65}{27} + \frac{2\zeta_2}{3} \right)+ \frac{875}{54} + \frac{16\zeta_2}{3} - \frac{20\zeta_3}{3} \right] .
\end{equation}
 The $\mathcal{Z}$-factor for massive quark, i.e. top quark, starts at $\alpha_s$, and the one-loop coefficient $\mathcal{Z}^{(1)}_{[t]}$ is given by
\begin{align}
\mathcal{Z}_{[t]}^{(1)}&=C_F\lt\{ \fc{2}{\eps^2}+\fc{2L_\mu+1}{\eps}+L_\mu^2+L_\mu+4+\zeta_2+ \eps\lt[\fc{L_\mu^3}{3}+\fc{L_\mu^2}{2}+(4+\zeta_2)L_\mu+8+\fc{\zeta_2}{2}-\fc{2\zeta_3}{3} \rt]\rt. \nn \\
&+\lt.\eps^2\lt[\fc{L_\mu^4}{12}+\fc{L_\mu^3}{6}+\lt(2+\fc{\zeta_2}{2}\rt)L_\mu^2+\lt(8+\fc{\zeta_2}{2}-\fc{2}{3}\zeta_3 \rt)L_\mu +16+2\zeta_2-\fc{\zeta_3}{3}+\fc{9}{20}\zeta_2^2\rt]\rt\} \, ,
\end{align}
where $L_{\mu} = \ln\lt(\mu^2/m_t^2\rt) $. The two-loop coefficient $\mathcal{Z}^{(2)}_{[t]}$ can be split into two parts:
\begin{equation}
\zcal_{[t]}^{(2)} = \zcal_{[t]}^{(2),l} + \zcal_{[t]}^{(2),t} \,,
\end{equation}
where $\zcal_{[t]}^{(2),l}$ contains contributions from gluon loops and light-quark loops; $\zcal_{[t]}^{(2),t}$ denotes the contribution from a loop insertion of top quark itself, and they are given by
\begin{align}
\mathcal{Z}_{[t]}^{(2),l}&=C_F^2\fc{2}{\eps^4}+\fc{1}{\eps^3}\lt[ C_F^2\lt(4L_\mu+2\rt)-\fc{11}{2}C_FC_A+n_lC_F\rt] +\fc{1}{\eps^2}\lt[ C_F^2\bigg(4L_\mu^2+4L_\mu+\fc{17}{2} \rt.\nn \\
&+\lt.2\zeta_2\bigg)-C_FC_A\lt(\fc{11}{3}L_\mu-\fc{17}{9}+\zeta_2 \rt)+n_lC_F\lt( \fc{2}{3}L_\mu-\fc{2}{9}\rt) \rt] +\fc{1}{\eps}\lt\{ C_F^2\lt[ \fc{8}{3}L_\mu^3+4L_\mu^2 \rt.\rt. \nn \\
&+\lt.\lt. \lt(17+4\zeta_2\rt)L_\mu+\fc{83}{4}-4\zeta_2+\fc{32}{3}\zeta_3 \rt] + C_FC_A\bigg[\lt( \fc{67}{9}-2\zeta_2\rt)L_\mu+\fc{373}{108}+\fc{15}{2}\zeta_2\rt. \ \nn \\
&-\lt.15\zeta_3 \bigg] + n_lC_F\lt( -\fc{10}{9}L_\mu-\fc{5}{54}-\zeta_2\rt) \rt\} + C_F^2\lt[ \fc{4}{3}L_\mu^4+\fc{8}{3}L_\mu^3+\lt(17+4\zeta_2\rt)L_\mu^2 \rt. \nn \\
&+\lt. \lt( \fc{83}{2}-8\zeta_2+\fc{64}{3}\zeta_3\rt)L_\mu+ \fc{561}{8}+\fc{61}{2}\zeta_2-\fc{22}{3}\zeta_3 -48\ln2\zeta_2-\fc{77}{5}\zeta_2^2 \rt] \nn \\
&+C_FC_A\lt[ \fc{11}{9}L_\mu^3+\lt( \fc{167}{18}-2\zeta_2\rt)L_\mu^2+\lt( \fc{1165}{54}+\fc{56}{3}\zeta_2-30\zeta_3\rt)L_\mu +\fc{12877}{648} \rt. \ \nn \\
&\lt. +\fc{323}{18}\zeta_2+\fc{89}{9}\zeta_3+24\ln2\zeta_2-\fc{47}{5}\zeta_2^2 \rt] +n_lC_F\lt[ -\fc{2}{9}L_\mu^3-\fc{13}{9}L_\mu^2+\lt(-\fc{77}{27}-\fc{8}{3}\zeta_2\rt)L_\mu \rt. \nn \\
&-\lt. \fc{1541}{324}-\fc{37}{9}\zeta_2-\fc{26}{9}\zeta_3\rt]\,,\\
\zcal_{[t]}^{(2),t} &= C_F T_F \left[ \frac{2}{\epsilon^3} + \frac{1}{\epsilon^2} \left( \frac{4}{3} L_\mu + \frac{8}{9} \right) + \frac{1}{\epsilon} \left( \frac{4}{9} L_\mu - \frac{65}{27} - 2\zeta_2 \right)-\frac{4}{9} L_\mu^3 - \frac{2}{9} L_\mu^2 \right. \nonumber
\\
&\left.  - \left( \frac{274}{27} + \frac{16\zeta_2}{3} \right) L_\mu + \frac{5107}{162} - \frac{70\zeta_2}{9} - \frac{4\zeta_3}{9} \right] \,.
\end{align}
Finally, $\mathcal{Z}$-factors for gluons up to NNLO are given by
\begin{align}
\mathcal{Z}_{[g]}^{(1)}&=\lt\{-\frac{2}{3\epsilon}-\frac{2}{3}L_{\mu}+\eps\lt(-\fc{1}{3}L_\mu^2-\fc{\zeta_2}{3}\rt)+\eps^2\lt(-\fc{1}{9}L_\mu^3-\fc{\zeta_2}{3}L_\mu+\fc{2\zeta_3}{9}\rt)\rt\} \,, \nn \\
\mathcal{Z}_{[g]}^{(2)}&=\lt(\mathcal{Z}_{[g]}^{(1)}\rt)^2+\fc{4}{3\eps}\lt(n_h+n_l\rt) T_F \mathcal{Z}_{[g]}^{(1)}+\mathcal{Z}_{[g]}^{(2),t}\,. \nn
\end{align}
where $\zcal_{[g]}^{(2),t}$ denotes the contribution from a loop insertion of top quark,
\begin{align}
\zcal_{[g]}^{(2),t} &= C_A T_F \left( \frac{\mu^2}{m_t^2} \right)^{2\epsilon} \left[ \frac{2}{\epsilon^3} + \frac{34}{9\epsilon^2} - \frac{2}{\epsilon} \left( \frac{22}{9} L_\mu + \frac{64}{27} - \zeta_2 \right) \right. \nonumber
\\
&\left. + \frac{22}{9} L_\mu^2 + \frac{358}{27} + \frac{4\zeta_2}{3} - 4\zeta_3 \right] - C_F T_F \left( \frac{\mu^2}{m_t^2} \right)^{2\epsilon} \left( \frac{2}{\epsilon} + 15 \right) .
\end{align}

\bibliographystyle{JHEP}
\bibliography{references_inspire}

\end{document}